\documentclass[12pt]{article}
\pdfoutput=1
\usepackage[left=2cm,right=2cm,top=1.5cm,bottom=1.5cm]{geometry}
\usepackage{multicol}
\usepackage{amssymb}
\usepackage{float}
\usepackage[usenames, dvipsnames]{xcolor}
\usepackage{graphicx}
\usepackage[multidot]{grffile}
\usepackage{subcaption}
\usepackage[final]{pdfpages}
\usepackage{amssymb,amsmath}
\usepackage{mathtools} 
\usepackage{listings} 
\usepackage{enumitem} 
\usepackage[style=numeric]{biblatex}
\usepackage{setspace} 
\usepackage[unicode=true,
  linktocpage,
  linkbordercolor={0.5 0.5 1},
  citebordercolor={0.5 1 0.5},
  colorlinks=true,
  linkcolor=blue]{hyperref}
\makeatletter
\newcommand*{\addFileDependency}[1]{
  \typeout{(#1)}
  \@addtofilelist{#1}
  \IfFileExists{#1}{}{\typeout{No file #1.}}
}
\makeatother

\DeclarePairedDelimiter{\ceil}{\lceil}{\rceil}

\newcommand{\pd}[2]{\frac{\partial #1}{\partial #2}}

\newcommand{\Cdata}{C_{\textrm{data}}}
\newcommand{\Ctdata}{\Tilde{C}_{\textrm{data}}}
\newcommand{\Cmodel}{C_{\textrm{model}}}

\newcommand{\btheta}{\boldsymbol{\theta}}
\newcommand{\degC}{^{\circ} \text{C}}
\newcommand{\tth}{^{\text{th}}}

\newcommand{\mum}{\textmu m}

\newcommand{\revision}[1]{#1}

\newcommand{\specialcell}[2][c]{%
  \begin{tabular}[#1]{@{}c@{}}#2\end{tabular}}

\begin{document}

\title{Parameter identifiability and model selection for partial differential equation models of cell invasion}

\author{
Yue Liu$^{1*}$, Kevin Suh$^{2}$, Philip K. Maini$^{1}$, Daniel J. Cohen$^{2,3}$, Ruth E. Baker$^{1}$ \\
$^{1}$Mathematical Institute, University of Oxford\\
$^{2}$Department of Chemical and Biological Engineering, Princeton University\\
$^{3}$Department of Mechanical and Aerospace Engineering, Princeton University\\
$^*$ Corresponding author: yue.liu@maths.ox.ac.uk
}
\date{}

\maketitle

\begin{abstract}


When employing mechanistic models to study biological phenomena, practical parameter identifiability is important for making accurate predictions across wide range of unseen scenarios, as well as for understanding the underlying mechanisms.
In this work we use a profile likelihood approach to investigate parameter identifiability for four extensions of the Fisher--KPP model, given experimental data from a cell invasion assay. We show that more complicated models tend to be less identifiable, with parameter estimates being more sensitive to subtle differences in experimental procedures, and that they require more data to be practically identifiable.  As a result, we suggest that parameter identifiability should be considered alongside goodness-of-fit and model complexity as criteria for model selection. 

\medskip
Keywords: parameter identifiability, model selection, profile likelihood, cell invasion, reaction-diffusion

\end{abstract}

\section{Introduction}

Partial differential equation (PDE) models have been widely employed across many areas of biology as a means to understand  mechanisms driving observed behaviours, as well as for making predictions of future behaviours. The complexity of a typical biological system means that often it is not clear which mechanisms should be incorporated into a PDE model, to what detail should they be described, and what functional form should they take. As a result, there can be multiple models proposed to describe the same system. For example, a variety of growth models has been proposed and adapted to describe the growth of tumors, coral reefs, and microbial growth~\cite{gerlee2013ModelMuddleSearch,peleg2011MicrobialGrowthCurves,simpson2022ParameterIdentifiabilityModel,}, and their advantages and disadvantages have been extensively studied. To be able to both make accurate predictions for unseen scenarios and elucidate underlying mechanisms it is crucial to be able to conduct model selection.

Criteria for model selection have mostly focused on balancing model complexity with the ability of a model to accurately reproduce experimental observations. Typical examples are the Akaike Information Criterion (AIC) and Bayesian Information Criterion (BIC)~\cite{aho2014ModelSelectionEcologists,burnham2004MultimodelInferenceUnderstanding}.
In this work, we will show that attention should also be paid to the issue of parameter identifiability in the process of model selection, which is ignored by information criteria. Parameter identifiability refers to the extent to which the parameters of a model can be accurately estimated from available data. Non-identifiable models can give rise to misleading conclusions about the nature of the underlying mechanisms, as well as inaccurate predictions. As such, it is important to ensure that model selection considers identifiability.

In this paper, we will explore the connections between model selection and parameter identifiability through the use of \textit{in vitro} experimental models of collective cell invasion, and four related candidate PDE models to describe them. We will use profile likelihoods for identifiability analysis, and investigate the impact of model complexity, data resolution, and experimental design, on parameter identifiability.

\subsection{Parameter identifiability}

The issue of parameter identifiability revolves around the question of whether it is possible to use available data to accurately estimate model parameters that describe the underlying mechanisms driving the biological system described by the model, accounting for observational errors. This inquiry revolves around two primary notions of identifiability found in the literature.
The first notion is structural identifiability, which refers to the ability to uniquely determine the values of the model parameters given an infinite amount of data~\cite{wieland2021StructuralPracticalIdentifiability}. In the context of PDE models, this is equivalent to ensuring that distinct sets of parameter values do not yield identical solutions. While several methods exist for determining structural identifiability for ordinary differential equation (ODE) models~\cite{chis2011StructuralIdentifiabilitySystems,raue2014ComparisonApproachesParameter}, methods for PDE models are restricted to specific classes of models, such as age-structured PDEs~\cite{renardy2022StructuralIdentifiabilityAnalysis} or systems of linear reaction-advection-diffusion equations~\cite{browning2023SIAPDEs}.

This paper, however, focuses on the more useful notion of practical identifiability. A model is considered practically identifiable if the parameters can be confidently identified using available data~\cite{wieland2021StructuralPracticalIdentifiability}. Practical identifiability is a stronger condition than structural identifiability, and its focus on the available data makes it more relevant for real-world applications. The main distinction between the two notions is that structural identifiability can be formulated as a property of the PDE model itself, while practical identifiability is a property of the combination of the PDE model, the error model, and data~\cite{raue2014ComparisonApproachesParameter}. In this paper, we will define a parameter to be practically identifiable if the 95\% confidence region obtained using profile likelihoods for this parameter is finite, and the profile likelihood function is unimodal. Moreover, for parameters that satisfies the above definition, we call a parameter more (or less) identifiable if its confidence interval is narrower (or wider) than another parameter.

Parameter identifiability directly affects the reliability of model predictions and the ability of a mechanistic model to precisely pin down biological mechanisms, since non-identifiability may lead the modeller to place confidence in a set of parameter values that, while able to produce model solutions close to experimental data under the conditions in which the experiments were conducted, may lead to model predictions that diverge from the behaviour of the biological system under a different set of conditions.

As practical parameter identifiability is dependent on both the model and data, our investigation will focus on both of these aspects. From a data-oriented perspective, we will investigate how the quality and quantity of data impacts parameter identifiability. Specifically, we will explore the extent to which experimental design, data analysis and processing impact parameter identifiability. The results from these investigations will provide guidance for the design of experiments, as well as for data collection and processing procedures, for the purpose of improving identifiability. We will also examine the relationship between data resolution and parameter identifiability.


\subsection{Model selection}

When we build a model for a given biological system, it is often difficult to determine the appropriate level of complexity, more specifically, which mechanisms to include, and to what level of detail, and which mechanisms to simplify or neglect. 
When the purpose of the model is to investigate a particular mechanism, it is sensible to choose a model that revolves around that mechanism~\cite{browning2022GeometricAnalysisEnables}, which provides a lower bound on complexity. Guidelines from~\cite{gerlee2013ModelMuddleSearch} argue that, while phenomenological models are acceptable for making predictions, a model built for the purpose of understanding the biological system should focus on mechanisms that can be concretely derived from the biology. This provides an upper bound on complexity, but in between there remains room for choice.

Increasing the level of complexity of a model by either including additional mechanisms, or using more parameters to fine-tune the description of the existing mechanisms, will potentially result in a model that is capable of fitting the given data more accurately. The downside of a more complicated model, besides over-fitting and making analysis more difficult, is that the model parameters often become less identifiable, since a change in one parameter can be more easily compensated for by changes in other parameters. This trade-off between complexity and identifiability leads us to consider the problem of model selection, where we aim to choose the most appropriate model from a collection of models of varying complexity.

Traditional tools used for addressing the dilemma between goodness-of-fit and model complexity include the AIC~\cite[Ch.~3.4]{konishi2008InformationCriteriaStatistical} and BIC~\cite[Ch.~9]{konishi2008InformationCriteriaStatistical}. These information criteria assign a score to models that reward goodness-of-fit, while penalising model complexity.  In this paper, we instead focus on using parameter identifiability as a tool for model selection, exploring the relation between complexity and identifiability in further detail.  We show that the model selected using information criteria, such as AIC or BIC, might contain parameters that are harder to identify than parameters in competing models that score less well, \revision{despite the wide-spread usage of these information criteria for model selection in mathematical biology, such as in cell invasion~\cite{warne2019UsingExperimentalData}, HIV infection~\cite{bortz2006ModelSelectionMixedEffects}, and ecology~\cite{aho2014ModelSelectionEcologists}}. We propose that identifiability analysis should be performed prior to model selection, and that models that are identifiable given the data available should be favoured.


\subsection{Identifiability in the context of cell invasion}

Collective cell invasion plays a central role in development, wound healing, and cancer, and there has been considerable interest in understanding and quantifying the mechanisms that drive, promote or hinder this phenomenon. 
In this work, we will use \textit{in vitro} collective cell invasion as a prototypical biological system to explore the ways in which identifiability issues can arise, and how the profile likelihood method can be used to analyse them. The data we will use come from a suite of barrier assay experiments, and we will use the Fisher equation and its generalisations as candidate models.

There is a long line of research on model fitting and parameter identification for cell invasion that leads up to this work.
In an early work, Sherratt and Murray (1990)~\cite{sherratt1990ModelsEpidermalWounda} used the Fisher model, and its generalisation, the Porous Fisher model, to describe wound healing from a qualitative perspective. These models were fitted to data from cell proliferation assays by Sengers et al.~(2007)~\cite{sengers2007ExperimentalCharacterizationComputational}. Later, Treloar and Simpson (2013)~\cite{treloar2013SensitivityEdgeDetection} and Simpson et al~(2013)~\cite{simpson2013QuantifyingRolesCell} showed that the Fisher model can be fitted to cell invasion data, using only the information on the location of the edge of the cell population. In contrast, Jin et al.~(2016)~\cite{jin2016ReproducibilityScratchAssays} fitted the Fisher and Porous Fisher models to cell invasion data consisting of the cell density profile. The paper employed the maximum likelihood method for parameter estimation, and goodness-of-fit as the main criterion for model selection, but it did not address parameter identifiability directly. In a later work, Vittadello et al.~(2018)~\cite{vittadello2018MathematicalModelsCell} used information on cell cycle dynamics, in addition to cell density data, to estimate model parameters. The same data were used by Simpson et al.~(2020)~\cite{simpson2020PracticalParameterIdentifiabilitya} to explore parameter identifiability by quantifying the uncertainty in parameter estimates using both profile likelihoods and Bayesian inference.

The method we use for parameter inference, profile likelihoods, has been used in a variety of studies in identifiability analysis. It is built upon the maximum likelihood estimator (MLE), and has often been compared to Markov Chain Monte Carlo (MCMC)-based methods for Bayesian inference. MCMC methods provide more information on parameter values, but are more expensive to compute. Raue et al.~(2009)~\cite{raue2009StructuralPracticalIdentifiability} discussed the application of profile likelihoods to detect non-identifiable parameters, illustrated with an  example in the context of ODE models.
This work continued in Raue et al.~(2013)~\cite{raue2013JoiningForcesBayesian}, which suggests a method that combines MCMC methods and profile likelihoods to inform data collection and iteratively refine parameter estimates.
A comparison between the methods of MCMC and profile likelihoods was carried out in~\cite{villaverde2022AssessmentPredictionUncertainty} for ODE models of varying complexity.
The application of profile likelihoods in model selection was discussed in Simpson et al.~(2022)~\cite{simpson2022ParameterIdentifiabilityModel}, which used three different ODE models to describe coral growth, and discussed the importance of parameter identifiability in model selection. Model selection for cell invasion models has been discussed by Warne et al. ~(2019) \cite{warne2019UsingExperimentalData}, who used Bayesian methods for model fitting, and information criteria for model selection. The authors emphasized the importance of model complexity as a factor in model selection, and identified BIC as the best overall criterion. However, the authors realised that information criteria cannot account for all aspects of model comparison. 


\subsection{Outline}

The contribution of this paper is the investigation of parameter identifiability for multiple PDE models using profile likelihoods, and a discussion as to the implications of practical identifiability in model selection. We also investigate the effects of data resolution and experimental design on parameter identifiability. Lastly, we demonstrate a link between parameter identifiability and model robustness in terms of consistency of parameter estimates, which motivates a mixed effects view of the system. Under this view, each experimental replicate can be seen as taking a sample of the parameter values from a certain distribution, which tends to be more dispersed for less identifiable models. The rest of the paper proceeds as follows. We describe the experimental procedures for the barrier assay, as well as data processing procedures, in Section~\ref{sec:experiment}. We introduce the suite of mathematical models, as well as the profile likelihood method for parameter inference and identifiability analysis, in Section~\ref{sec:math_methods}. The results are presented in Section~\ref{sec:results}, and we discuss their significance with respect to model selection and experimental design in Section~\ref{sec:discussion}.
    
    
\section{Experimental methods and data}\label{sec:experiment}

\begin{figure}[htbp]
    \centering
    \begin{subfigure}[h]{0.2\textwidth}
        \centering
        \caption{}
        \includegraphics[width=\textwidth]{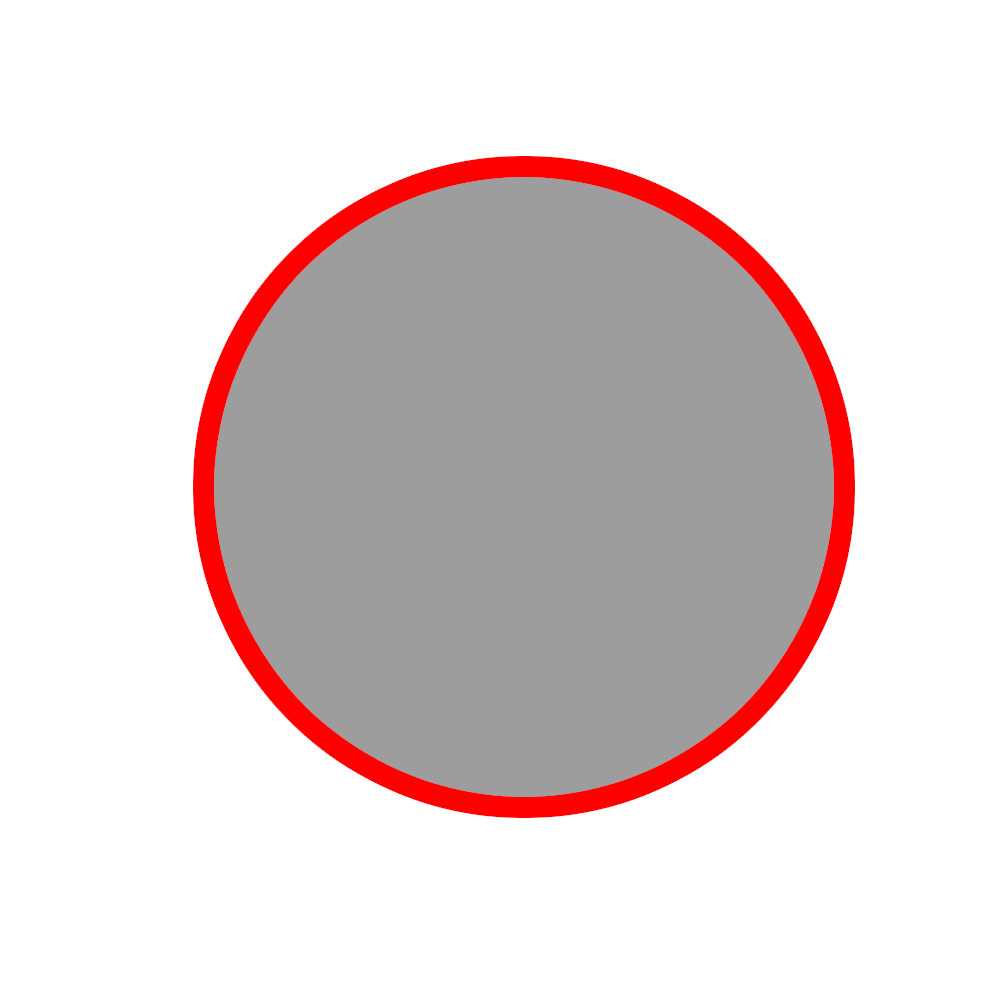}
        \includegraphics[width=\textwidth]{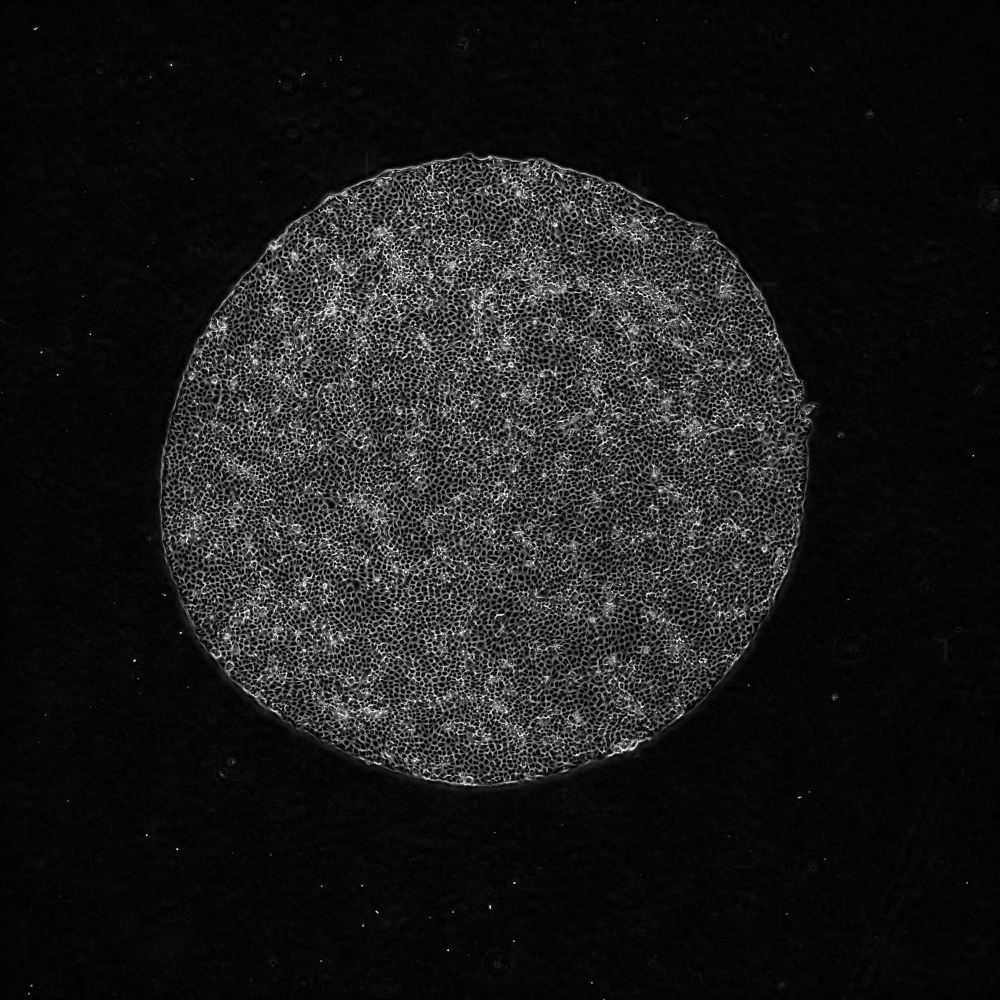}
        \includegraphics[width=\textwidth]{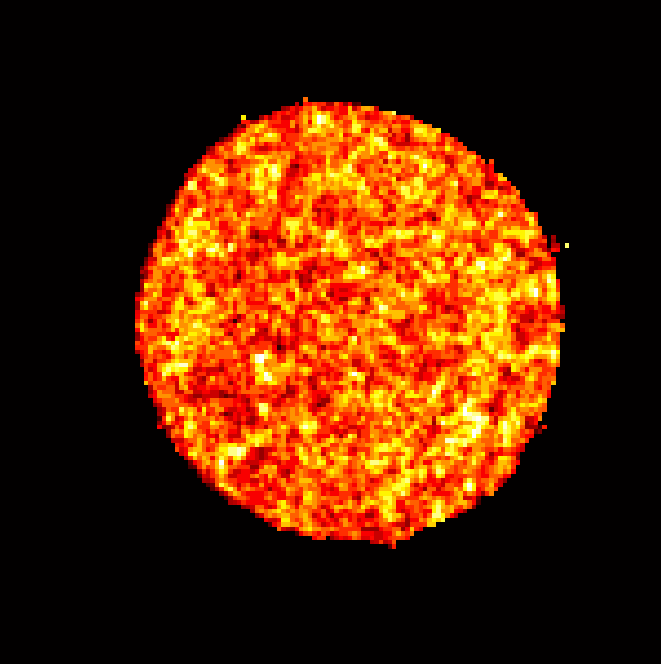}
    \end{subfigure}
    \begin{subfigure}[h]{0.2\textwidth}
        \centering
        \caption{}
        \includegraphics[width=\textwidth]{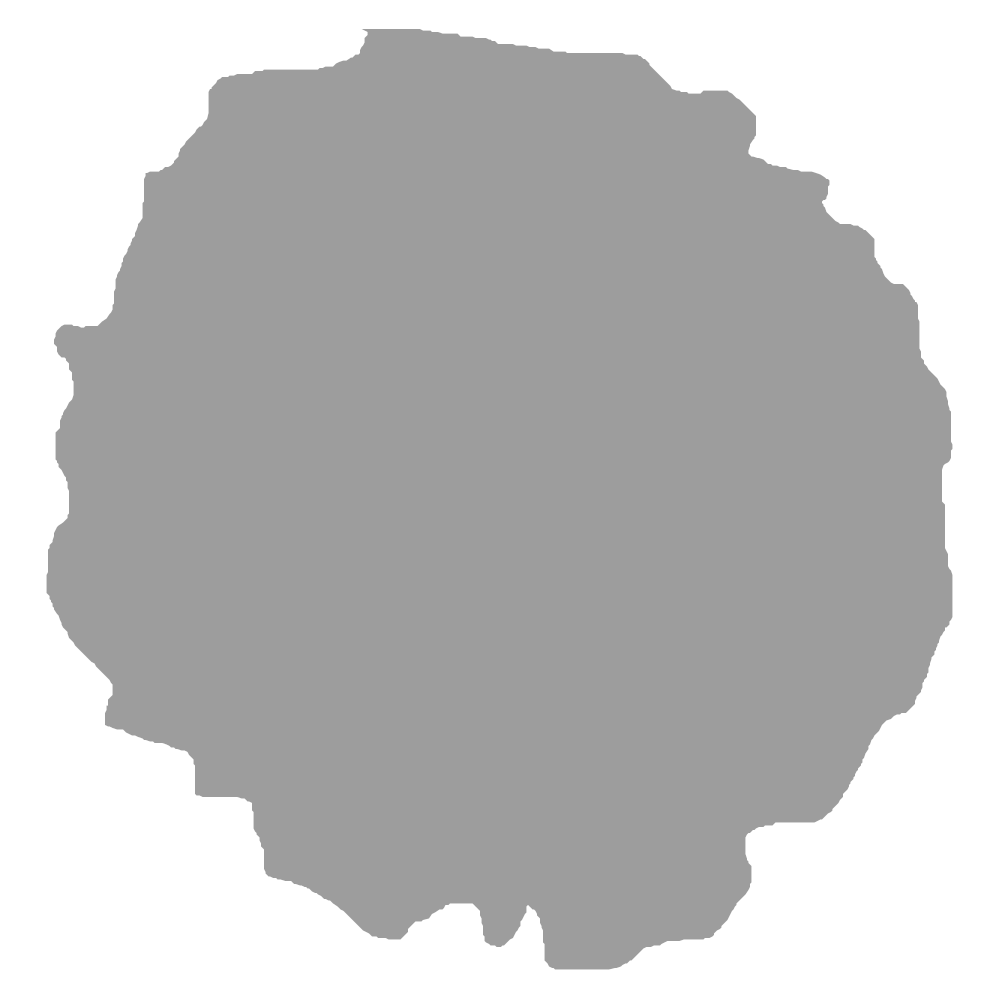}
        \includegraphics[width=\textwidth]{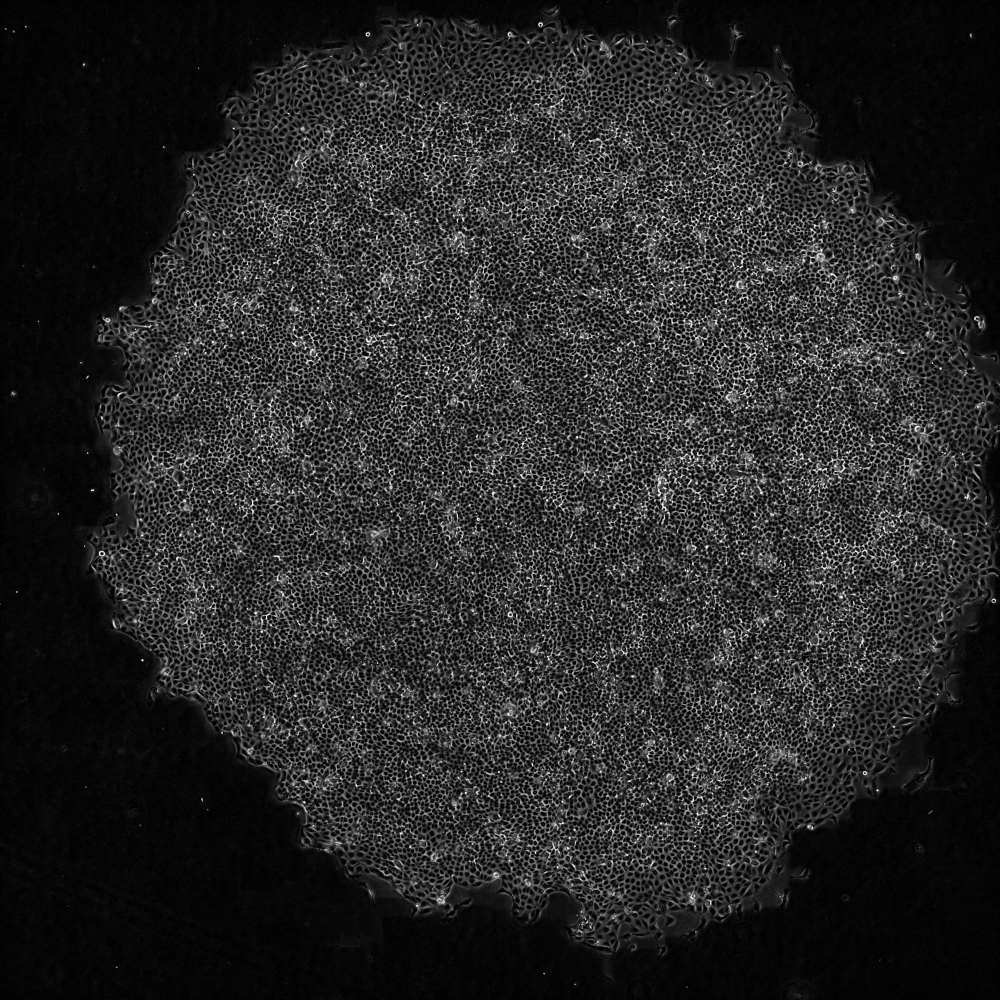}
        \includegraphics[width=\textwidth]{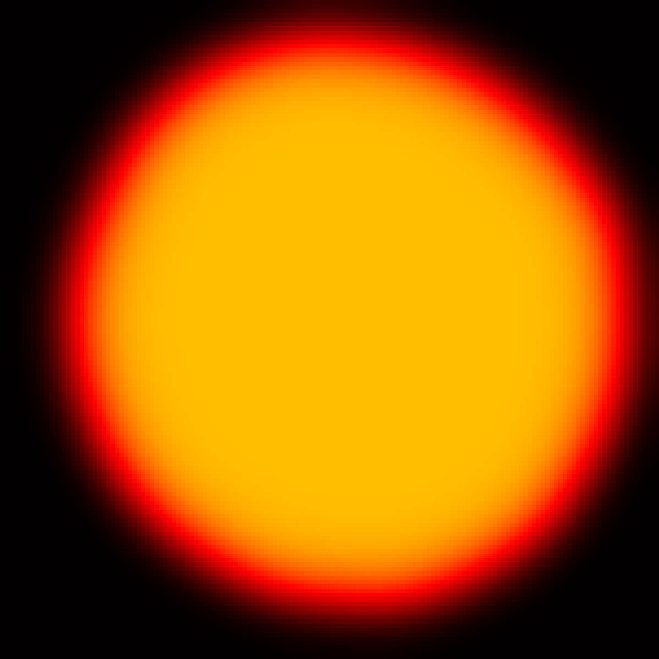}
    \end{subfigure}
    \begin{subfigure}[h]{0.2\textwidth}
        \centering
        \caption{}
        \includegraphics[width=\textwidth]{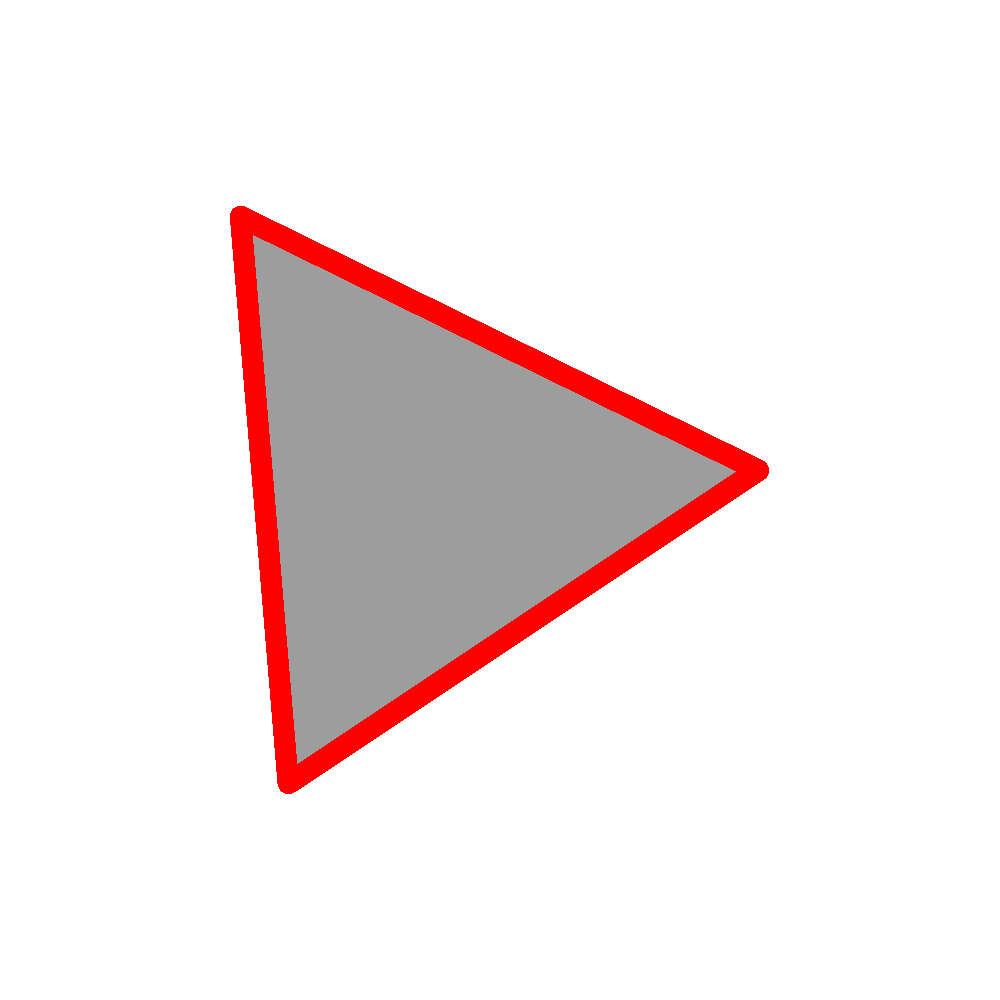}
        \includegraphics[width=\textwidth]{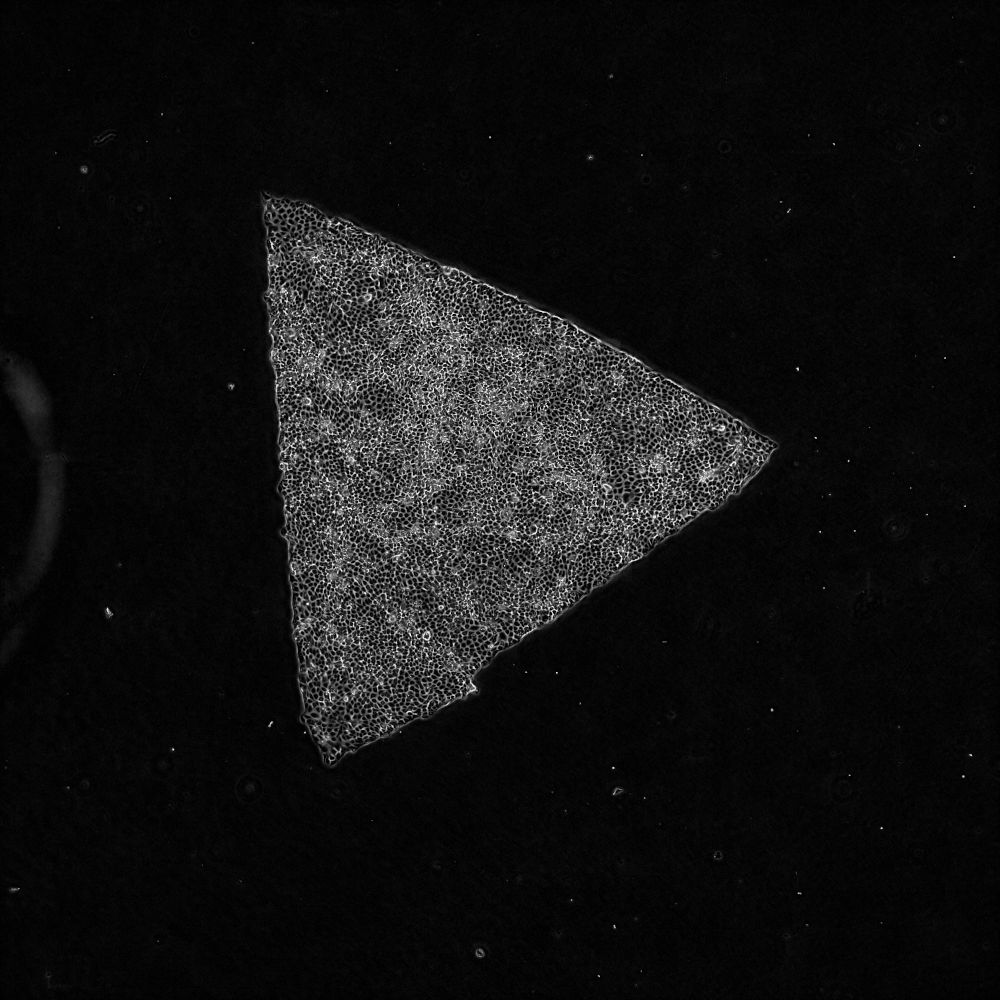}
        \includegraphics[width=\textwidth]{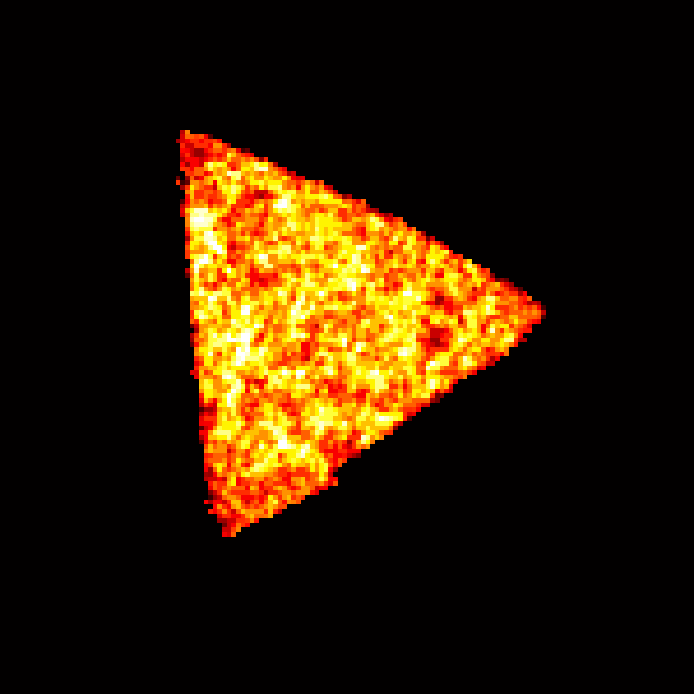}
    \end{subfigure}
    \begin{subfigure}[h]{0.2\textwidth}
        \centering
        \caption{}
        \includegraphics[width=\textwidth]{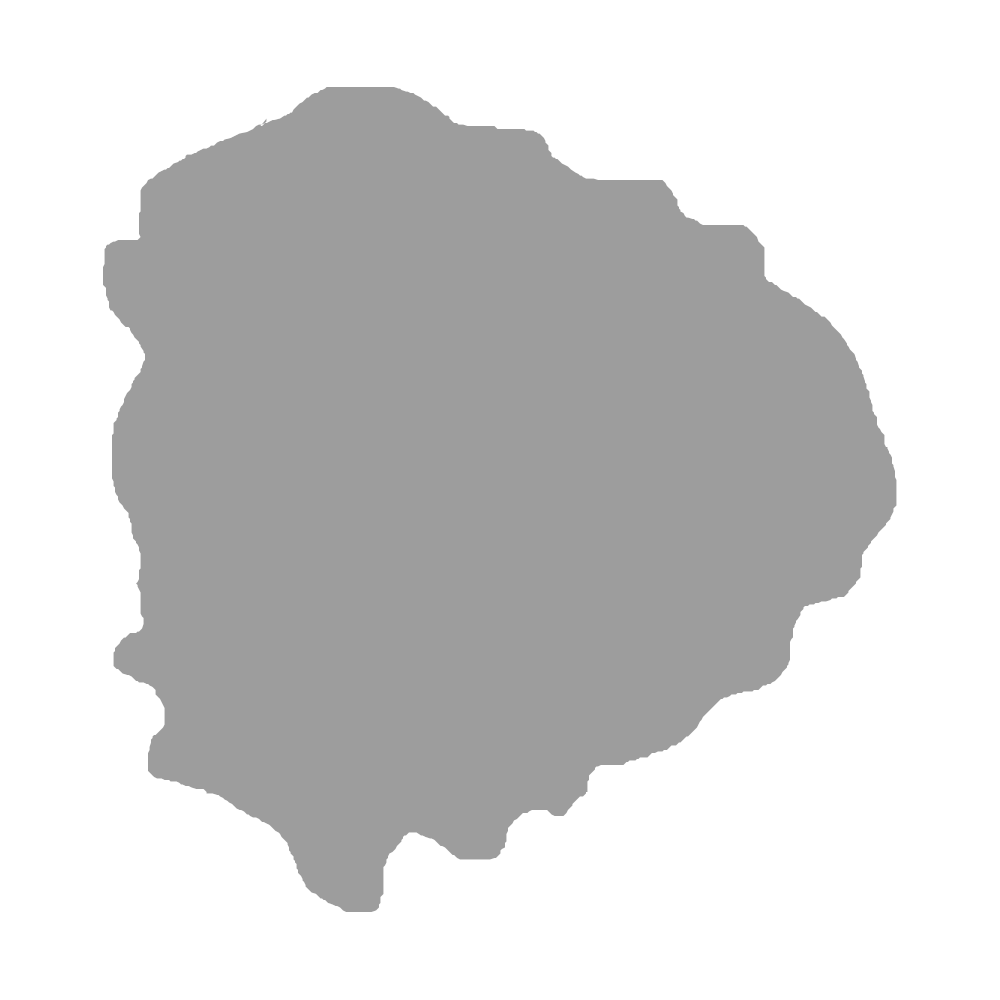}
        \includegraphics[width=\textwidth]{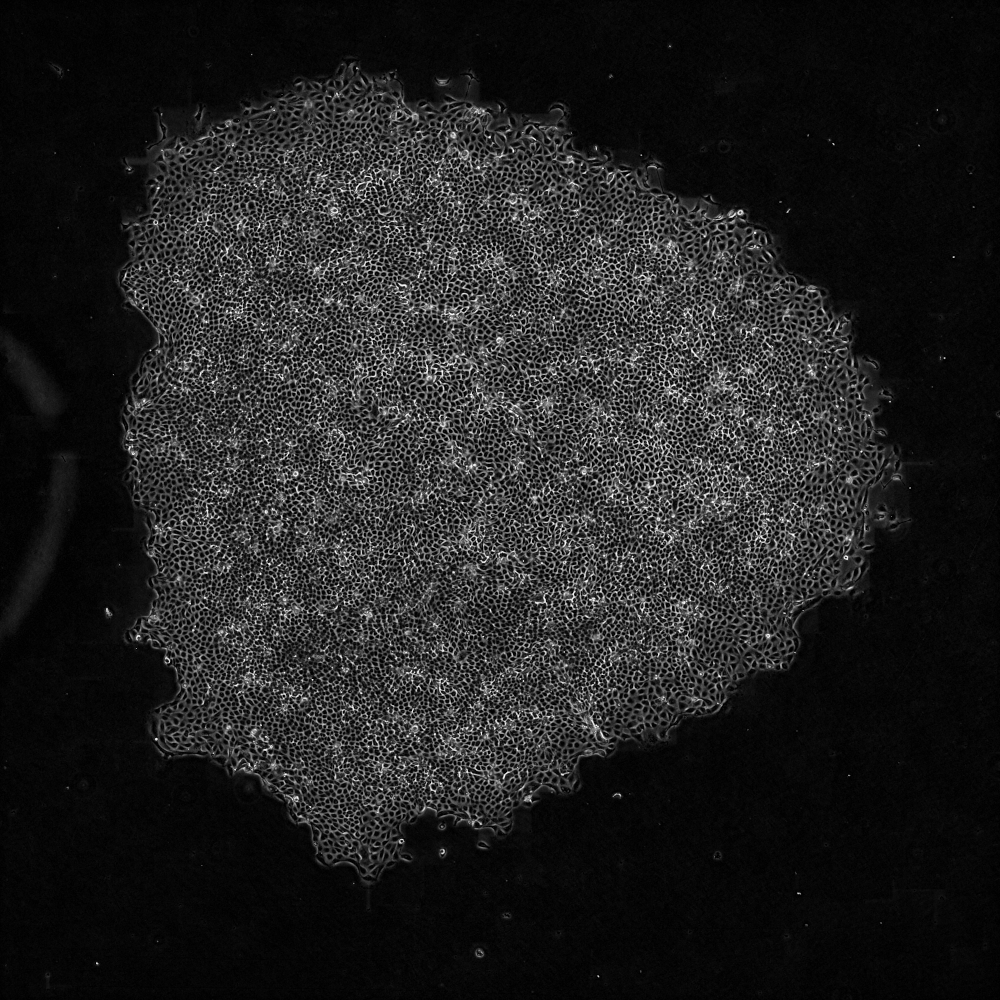}
        \includegraphics[width=\textwidth]{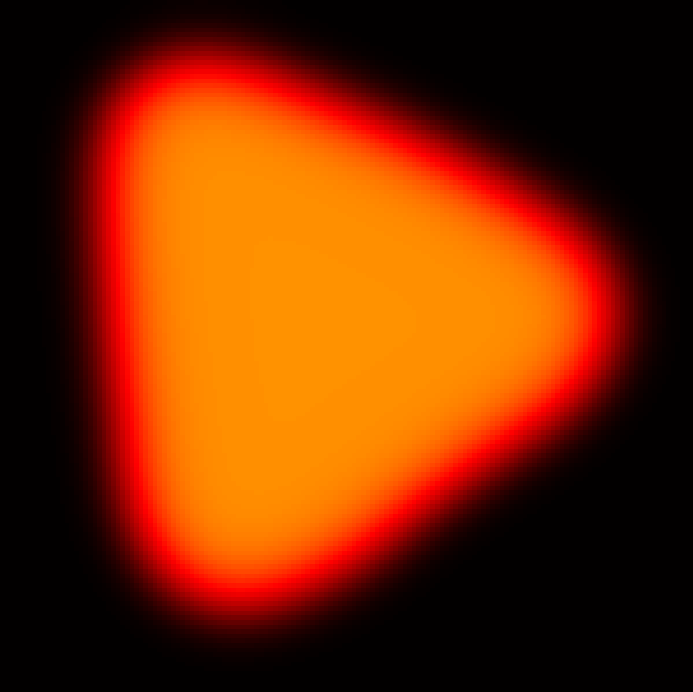}
    \end{subfigure}
    \caption[Illustration of the barrier assay]{Illustration of various stages of the barrier assay. In the first row are simplified representations of the cell culture, where the grey areas represent the regions covered by cells, and the red lines represent the barriers. The second row contains images taken from the experiments. 
    The field of view of the camera is limited to a square region of size $4380$ \mum\ by $4380$ \mum\ in the centre of the plate, while the plate itself extends further beyond. Since the cells tend to stay well within the field of view, except toward the very end of the experiment, edge effects are not important.
    The plate is sufficiently large such that the cells never reach the edge in any experiments.
    The third row represents snapshots from simulations of the Standard Fisher model (see Table~\ref{tab:models}) with the free parameters set to the maximum likelihood estimator (MLE) computed from the corresponding experiments. The colours represents cell density, with black indicating a complete absence of cells, while the cell density increases as the colour progresses from red to light yellow.
    (a) Initial condition of Experiment 1, where the initial confinement region is circular. The barrier is removed at $t=0$ h. (b) The state of the experiment at $t=20$ h. The cells have spread out to cover most of the visible domain. 
    (c-d) Shows the initial conditions at $t=0$ h and a later stage $t=20$ h of Experiment 5. The experiment is similar to Experiment 1, except with a triangular barrier.
    }
    \label{fig:exp_data}
\end{figure}

Barrier assays, also known as tissue expansion assays, are a way to observe cell invasion \textit{in vitro}. Briefly speaking, a barrier assay involves preparing a plate with a barrier that encloses a central region which cells cannot move in or out of. 
Cells are planted within the central region, and then given sufficient time to proliferate to form a collective monolayer within the central region. The barrier is then removed, allowing the cells to propagate outward, and invade the rest of the plate, which is initially absent of cells. For this work, eight experiments were carried out in total, with the first four (Experiment 1, 2, 3, 4) using a circular barrier, and the later four (Experiment 5, 6, 7, 8) using a triangular barrier.  The experiments were otherwise identical.
Selected images from the experiments are shown in Fig.~\ref{fig:exp_data}, along with corresponding schematic representations of the experiment, and snapshots of model solutions at the corresponding time point. The detailed experimental protocols are provided in Sections~\ref{sec:cells} and~\ref{sec:barrier}.


\subsection{Cell culture}\label{sec:cells}

Madin-Darby Canine Kidney (MDCK-II) cells are cultured in low-glucose Dulbecco’s modified Eagle medium (DMEM) with 10\% FBS and 1\% Penicillin-Streptomycin. Cells were split every 1 to 2 days, depending on the conditions of the cells. MDCK cells were treated with TrypLE for 8 minutes to be detached from the TC plastic dish. Detached cells were 1:2 diluted with the culture media and centrifuged at $300$ g at RT for 3 minutes. Cell pellets were then re-suspended with culture media and seeded in the new dish.


\subsection{Barrier assay}\label{sec:barrier}

The tissue-culture treated plastic dish is coated with $50$ \textmu g/mL Collagen 4 in PBS at $37\degC$ for 30 min. 
A $250$ \mum\ thick PDMS (polydimethylsiloxane) membrane cut with a Silhouette Cameo vinyl cutter was deposited on the TC (tissue-culture treated) plastic dish, which confines the cells to the initial region.
A $2.25 \times 10^6$ cells/mL MDCK cell suspension was seeded in each stencil. The seeding volume was $0.44$ \textmu L per mm$^2$ of tissue area. The TC plastic dish was incubated at $37\degC$ for 30 minutes to induce cell attachment to the collagen surface. Subsequently, the dish was flooded with the culture medium, and cells were incubated for 19 hours to form a confluent monolayer. Stencils were removed from the dish with tweezers at the start of the barrier assay. The expansion of tissue monolayer was imaged from this point onward every 20 minutes for 24 hours with $4\times$ phase objective.


\subsection{Image analysis}

From the $4\times$ phase images, nuclei locations were reconstructed using a tool based on convolutional neural networks \cite{lachance2020PracticalFluorescenceReconstruction}. The local cell density was calculated by counting the number of nuclei centroids in each density measuring box. The box size is $58.4$ \mum $ \times 58.4$ \mum, with 50\% overlap between neighboring boxes.


\section{Mathematical methods}\label{sec:math_methods}


\subsection{PDE models of cell invasion}

To describe the cell invasion process observed in the barrier assay experiments, we will use the Fisher--KPP equation, and its extensions, to model the evolution of cell density over time. The original Fisher model was proposed in~\cite{fisher1937WaveAdvanceAdvantageous}, and studied in a more general context in \cite{kpp}. More recently in cell biology, this model and its generalisations have been used to model a diverse range of phenomena including wound healing~\cite{habbal2014AssessingAbility2D,maini2004TravellingWavesWound} and tumour growth~\cite{flegg2020CurrentPerspectiveWound,jackson2015PatientSpecificMathematicalNeuroOncology,shyntar2022TumorInvasionParadox}.

We summarise the Fisher model and its generalisations as
\begin{subequations}
\begin{align}
\pd{C}{t} &= \nabla \cdot (D(C) \nabla C) + f(C),\\
D(C) &= D_0 \left(\frac{C}{K} \right)^\eta, \\
f(C) &= rC^\alpha \left[ 1-\left(\frac{C}{K} \right)^\gamma \right]^\beta, \\
&0 < x,y < L, t > 0. \notag
\end{align}\label{eqn:model}
\end{subequations}
In the context of our application, $C(x,y,t) \geq 0$ represents cell density, $D(C) \geq 0$ is the diffusion coefficient, and the function $f(C)$ represents net cell proliferation, with $K>0$ representing the carrying capacity. The parameters $D_0>0$ and $r>0$ scale the rates of diffusion and proliferation, respectively, and $\alpha, \beta, \gamma, \eta$ are non-negative ``shape parameters" that allow fine-tuning of $D(C)$ and $f(C)$. We impose zero-flux boundary conditions at $x = 0, L$ and $y=0,L$. 

We consider four different models. The original Fisher model takes $D(C)=D_0$, and $f(C)=rC(1-C/K)$, which corresponds to fixing $\alpha=\beta=\gamma=1$, and $\eta=0$. The Porous Fisher model generalises the diffusion term by allowing $\eta \geq 0$ to be free. This model has been used to describe cell invasion and wound healing~\cite{falco2023QuantifyingTissueGrowth,jin2016ReproducibilityScratchAssays,mccue2019HoleclosingModelReveals,sengers2007ExperimentalCharacterizationComputational,sherratt1990ModelsEpidermalWounda,}. The intuitive motivation is that as cell density increases, a crowding effect will push the cells to move into less dense areas more quickly, leading to density-dependent diffusion. The theoretical properties of the model have been studied in many works, including~\cite{pablo1998TravellingWaveBehaviour,witelski1995MergingTravelingWaves}. The Richards model, proposed in~\cite{richards1959FlexibleGrowthFunction}, and the Generalised Fisher models both generalise the proliferation term by allowing $\gamma \geq 0$ and $\alpha, \beta \geq 0$ to assume values other than unity, respectively. The Richards and Generalised Fisher models, among other growth laws, were analysed in the context of ODE models in~\cite{tsoularis2002AnalysisLogisticGrowth}. We summarise the four models in Table~\ref{tab:models}, and the numerical methods used for simulating these models are given in Supplementary Material~\ref{apx:fdm}.


\begin{table}[H]
    \centering
    \begin{tabular}{c||c|c}
        Model name & Free parameters & Fixed parameters \\
        \hline\hline
        Standard Fisher & $D_0, r, K$ & $\alpha=\beta=\gamma=1, \eta=0$ \\
        \hline
        Porous Fisher & $D_0, r, \eta, K$ & $\alpha=\beta=\gamma=1$ \\
        \hline
        Richards & $D_0, r, \gamma, K$ & $\alpha=\beta=1, \eta=0$ \\
        \hline
        Generalised Fisher & $D_0, r, \alpha, \beta, K$ & $\gamma=1,\eta=0$  \\
    \end{tabular}
    \caption{Summary of models considered, as special cases of Eq.~\eqref{eqn:model}.}
    \label{tab:models}
\end{table}


\subsection{Parameter identifiability with profile likelihoods}

For a given model, let $P$ be the number of free parameters, $\btheta$ denote the vector of free parameters in the model, and $\btheta_{-i}$ denote the parameter vector with $\theta_i$, the $i\tth$ parameter, removed\footnote{We may also replace the subscript with the name of the replaced parameter for readability.}.  For the Standard Fisher model, for example, $P=3$ and $\btheta=(D_0, r, K)$, and $\btheta_{-2}=\btheta_{-r}=(D_0, K)$. Let $\Cdata(x,y,t)$ be the cell density measurements, and let $\Cmodel(x,y,t; \btheta)$ be the solution to Eq.~\eqref{eqn:model} with parameter $\btheta$. For simplicity, we assume that the observed cell density is generated by the model with given model parameter set $\Hat{\btheta}$, and perturbed by independent and identically distributed (i.i.d.)~observation noise at each data point that follows a Gaussian distribution with zero mean and unknown variance $\sigma^2$. That is,

\begin{equation}
\Cdata(x_i,y_j,t_k) = \Cmodel(x_i,y_j,t_k; \Hat{\btheta}) + \epsilon_{ijk}, \quad \epsilon_{ijk} \sim \mathcal{N}(0,\sigma^2).
\label{eqn:obs_error}
\end{equation}
Let $L(\Cdata | \btheta,\sigma)$ be the likelihood function. Let $\btheta^*, \sigma^*$ be the maximum likelihood estimator (MLE).
We define the normalised profile log-likelihood function $l_i(\theta_i')$ for parameter $\theta_i$ to be
\begin{equation}
l_i(\theta_i') = \max_{\btheta_{-i}, \sigma} \left[ \log L(\Cdata | \btheta,\sigma)|_{\theta_i=\theta_i'} \right] - \log L(\Cdata | \btheta^*, \sigma),
\label{eqn:log-profile-likelihood}
\end{equation}
which will be referred to simply as the profile likelihood function for brevity. We can similarly define a bi-variate profile likelihood, $l_{i,j}$, as 
\[l_{i,j}(\theta_i',\theta_j') = \max_{\btheta_{-i,-j}, \sigma} \left[ \log L(\Cdata|\btheta,\sigma) |_{\theta_{i}=\theta_i', \theta_j=\theta_j'} \right] - \log L(\Cdata|\btheta^*,\sigma^*).\] 

Following \cite{murphy2000ProfileLikelihood} and \cite{simpson2022ParameterIdentifiabilityModel}, we use the profile likelihood function to define an approximate 95\% confidence interval for a single variable as
\[\{\theta_{i} | l_{i}(\theta_i) > -\chi^2(0.95; 1)/2 \approx -1.92\},\]
and a joint confidence region for two variables as
\[\{\theta_{i},\theta_{j} | l_{i,j}(\theta_i, \theta_j) > -\chi^2(0.95; 2)/2 \approx -3.00\},\]
where $\chi^2(\cdot; m)$ denotes the inverse of the cumulative distribution function (cdf) of a $\chi^2$ distribution with $m$ degrees of freedom. The shape of the univariate profile likelihood curve indicates whether the parameter is practical identifiable. A profile likelihood curve that is unimodal, smooth, and decays quickly away from its peak (which occurs at the MLE) suggests that the parameter is identifiable. Non-identifiability can be reflected in a profile likelihood in many different ways, such as multi-modality, slow decay away from the MLE, or a flat top.


\section{Results}\label{sec:results}

We now present the results from the identifiability analysis of the four models using profile likelihoods.
A preliminary exploration with synthetic datasets, presented in Supplementary Material Section~\ref{apx:synthetic},  shows that profile likelihoods can recover the ground truth parameter values in the absence of model mis-specification.
In Section~\ref{sec:pl_one_dataset}, we show that all models considered are practically identifiable. Then, in Section~\ref{sec:full_vs_reduced}, we show that slight changes to data processing procedures have a major impact on the parameter estimates for more complicated models, but not for simple models. In Section~\ref{sec:consistency}, we compare the profile likelihoods across multiple experimental datasets to observe a relation between model complexity and consistency of parameter estimates. Finally, in Section~\ref{sec:data_resolution}, we investigate the relation between data resolution and practical identifiability.


\subsection{All four models are practically identifiable}\label{sec:pl_one_dataset}

\begin{figure}[htbp]
    \centering
    \includegraphics[width=0.95\columnwidth]{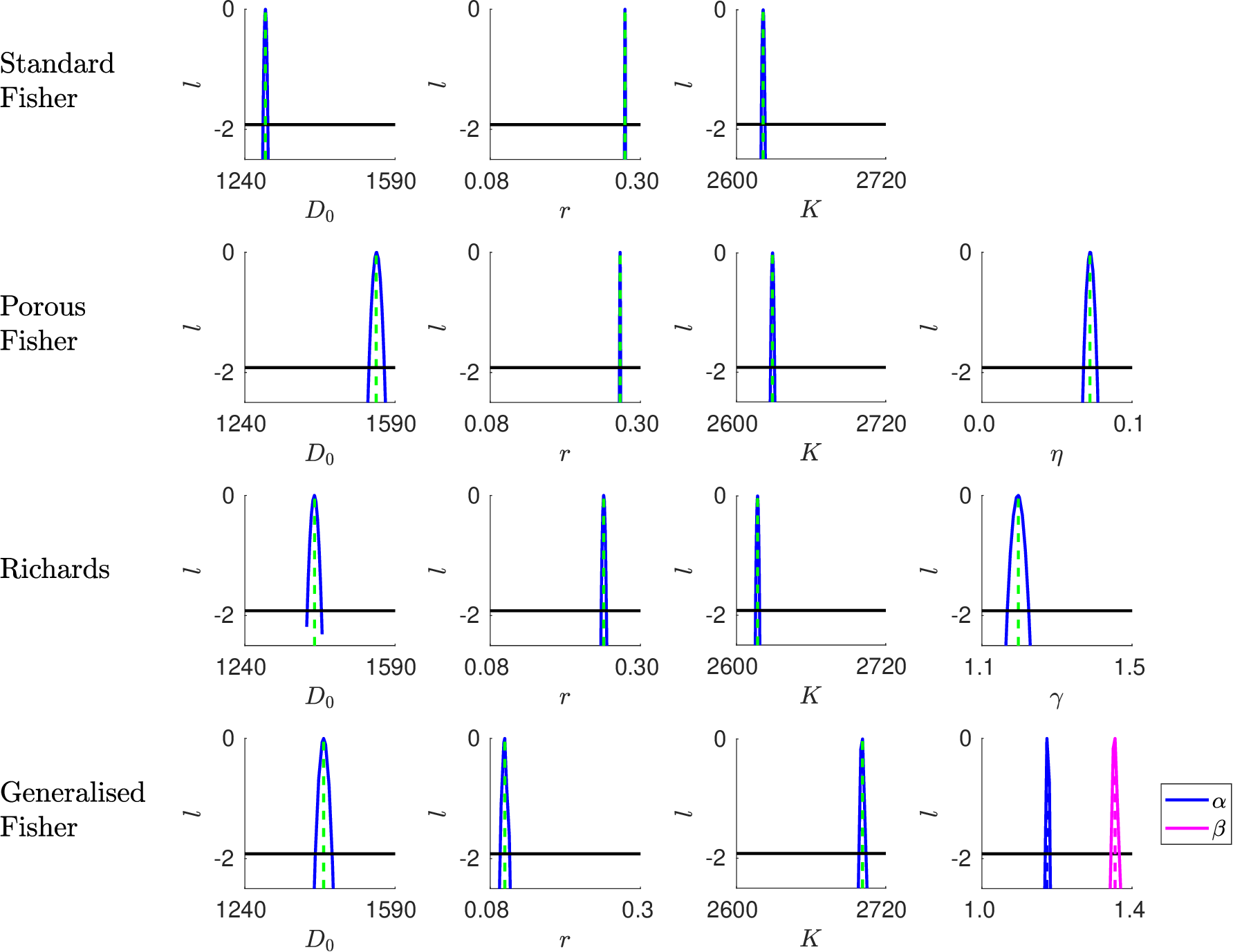}
    \caption[Profile likelihoods for Experiment 1, full dataset]{Profile likelihoods using data from Experiment 1. The vertical dashed lines mark the MLE for each parameter, and the black horizontal line at $-1.92$ marks the threshold for the 95\% confidence interval.}
    \label{fig:figtable_xy1_2d}
\end{figure}

First, we show that all four models are practically identifiable. We present the profile likelihoods for Experiment 1 in Fig.~\ref{fig:figtable_xy1_2d}, which is arbitrarily chosen as an illustrative representative for the purpose of this section. The results for the other experiments are qualitatively similar. We present the profile likelihoods for the other experiments, and the MLEs, the 95\% confidence intervals, and AIC and BIC for each model for each experiment, in Supplementary Material Section~\ref{apx:mle_tables}. Visual inspection indicates that the MLEs for all four models produce solutions that fit the data very well. As such, it is difficult to select between models by such inspection alone.


Observe that in Fig.~\ref{fig:figtable_xy1_2d}, all profile likelihood curves are unimodal, and roughly parabolic in shape, and the confidence intervals are narrow. These observations, and similar results from the other experiments (shown in Supplementary Material Section~\ref{apx:mle_tables}), suggest that all models are practically identifiable given available data from any one of the eight experiments. Comparing the Standard Fisher model with the more complicated models we notice that, as model complexity increases, the profile likelihood curves broaden, and therefore the confidence intervals widen, making the parameters less identifiable. This suggests that increases in model complexity tend to lead to decreases in parameter identifiability. Intuitively, this is because more complicated models have more freedom to compensate for a mis-identification in one parameter by adjusting the other parameters.


\subsection{Complicated models are more sensitive to data processing}\label{sec:full_vs_reduced}

We now turn to the question of whether the way we process the data impacts parameter identifiability.
In the case where the barrier is circular, the cell population  remains approximately radially symmetric throughout the experiment, as expected. In this case, we can average the data radially to obtain a reduced dataset $\Ctdata(\rho,t)$. It is tempting to work with this reduced dataset instead of the full dataset, since it makes model simulation, and therefore parameter inference and identifiability analysis, computationally much cheaper.

\begin{figure}[htbp]
    \centering
    \includegraphics[width=0.95\textwidth]{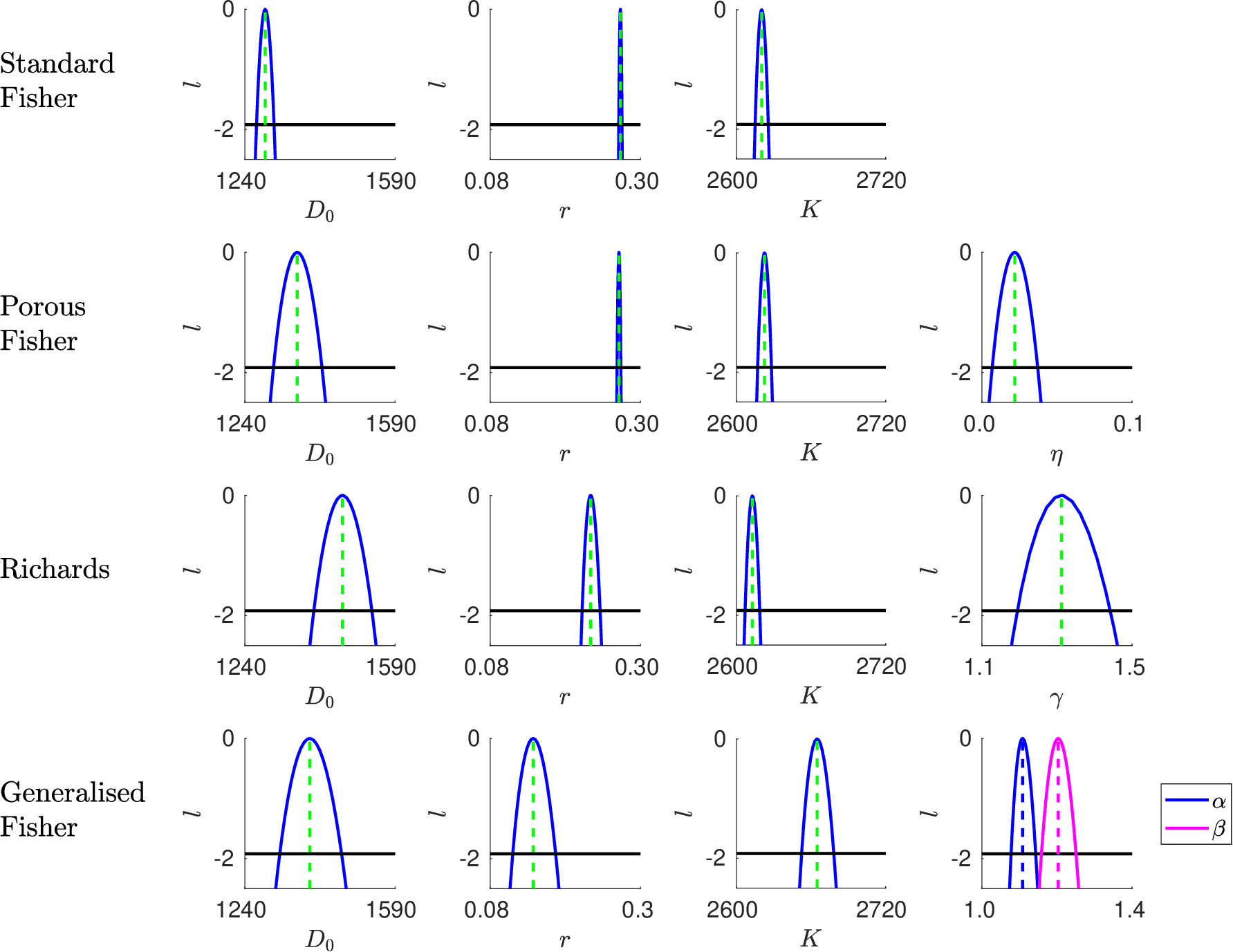}
    \caption[Profile likelihoods for Experiment 1, reduced dataset]{
    As per Fig.~\ref{fig:figtable_xy1_2d}, except using the radially averaged dataset $\Ctdata(\rho,t)$. The axes are the same as in Fig.~\ref{fig:figtable_xy1_2d}. Notice that the profile likelihood curves are broader compared to Fig.~\ref{fig:figtable_xy1_2d}, and the location of the MLE for the Porous Fisher, Richards, and Generalised Fisher models are shifted, but remain virtually unchanged for the Standard Fisher model. 
    }
    \label{fig:figtable_xy1_1d}
\end{figure}

In Fig.~\ref{fig:figtable_xy1_1d}, we show the profile likelihoods computed using the radially averaged dataset from Experiment 1.  Comparing Fig.~\ref{fig:figtable_xy1_1d} and Fig.~\ref{fig:figtable_xy1_2d}, we notice that, in general, the profile likelihood curves are broader for the radially averaged dataset compared to the full dataset. This is expected, since the full dataset has many more data points. A more surprising observation is that, while the MLEs for the parameters in the Standard Fisher model remain consistent between the two datasets, the MLEs for the other three models are much less consistent. 
These observations show that the more complicated models are more sensitive to the way the data are processed or represented.

In this instance, a likely explanation is that computing the radial average smooths the cell density near the edge of the population. The more complicated models are more sensitive to such subtle differences in the data, which stems from their increased flexibility to fit the data. Another way to look at this is that averaging reduces the observational noise present in the data, therefore preventing the models from over-fitting the data. The more complex models may have a higher propensity to over-fit, which explains why these models display a greater difference in the profile likelihoods between the two datasets.

This sensitivity of parameter estimation to data representation means care must be taken during data processing. If we wish to use a complicated model, then we should use the full density profile because it suffers less information loss. On the other hand, if we are content with a simple model like the Standard Fisher model, then it would be preferable to use a dataset of reduced dimension, because it yields the same result as the full dataset, but makes the computations cheaper.


\subsection{Consistency of parameter estimates across experimental replicates indicates practical identifiability}\label{sec:consistency}

\begin{figure}[htbp]
    \centering
    \includegraphics[width=0.95\columnwidth]{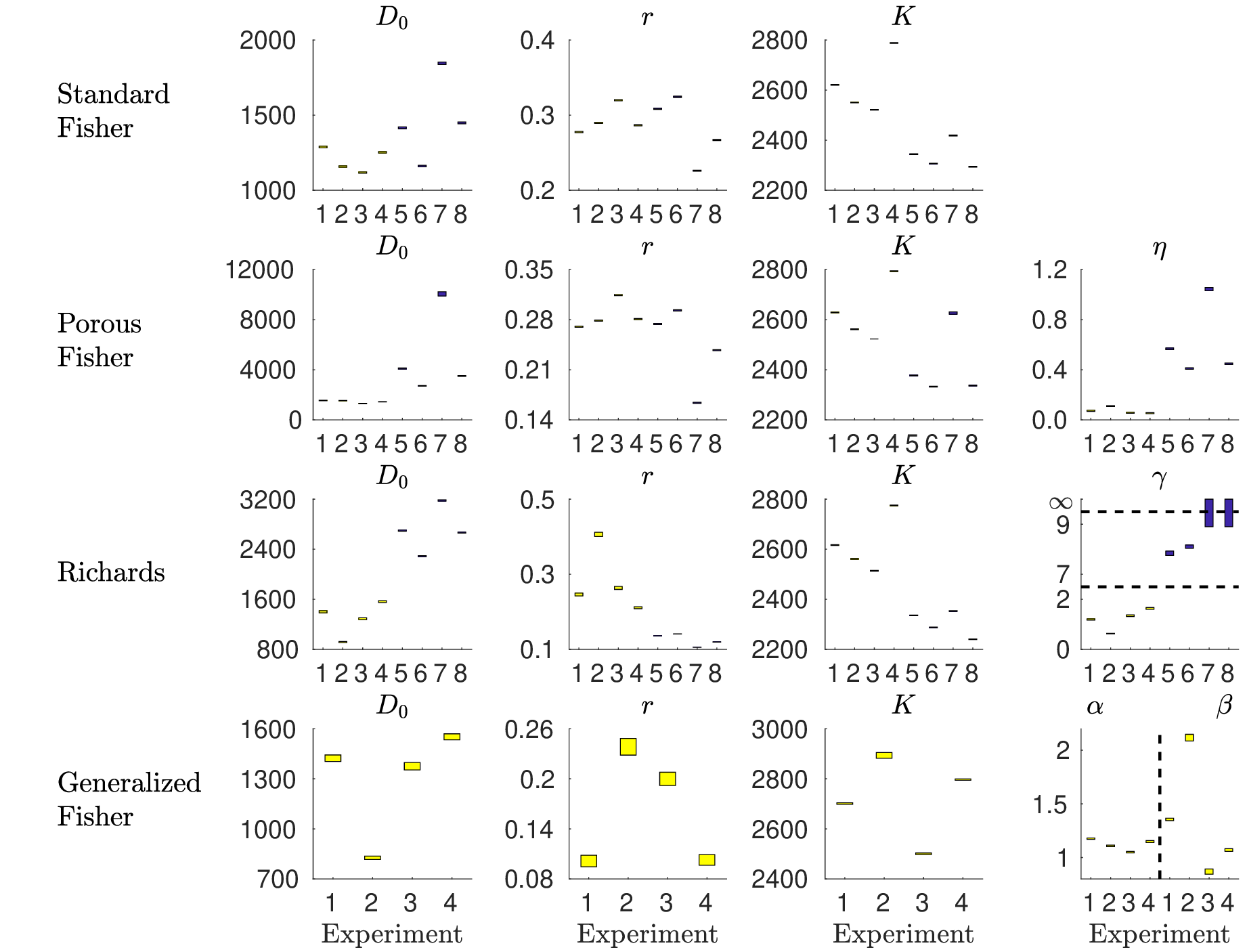}
    \caption[Comparison of confidence intervals for the parameters]{Comparison of the 95\% confidence intervals for the parameters in the four models obtained using profile likelihoods on the eight experimental datasets. The confidence intervals are represented as ranges by vertical bars. The experiments with circular initial conditions (Experiments 1-4) are labelled in yellow, while the ones with triangular initial conditions (Experiments 5-8) are labelled in blue. Note that some vertical axes do not start from zero. Note that the horizontal dashed lines in the figure for $\gamma$ in the Richards model indicates breaks in $y-$axis.
    }
    \label{fig:conf_interval_compare_all}
\end{figure}

By examining and comparing the profile likelihood results across all experimental replicates, we found that the parameter estimates are less consistent for the more complicated models compared to the simpler Standard Fisher model, in the sense that the variances of the eight MLEs of analogous parameters are higher for the more complicated models. In Fig.~\ref{fig:conf_interval_compare_all}, we present the confidence intervals for all parameters in the four models found across the eight experimental replicates, except for the Generalised Fisher model, where we only present the results from the first four experiments. \footnote{The profile likelihood computation for the other four experiments could not be completed in reasonable time, \revision{due to difficulties in performing the optimisations during the computations for the profile likelihoods. We did not pursue this technical challenge further since it is beyond the scope of the paper.}}

For all parameters, the widths of the confidence intervals for individual experiments are much narrower compared to the range of estimates across all experiments. For example, the widths of the confidence intervals for $D_0$ in the Standard Fisher model are on the order of $10$ \mum$^2$/h, whereas the MLEs for $D_0$ for Experiment 3 and 7 differ by around $700$ \mum$^2$/h. The confidence intervals for the same parameter estimated from different experiments rarely overlap; these variations in parameter estimates reflect the high variability commonly observed in biological processes.  This suggests that it would be appropriate to model the system using a mixed effects framework, where the parameter values corresponding to each experiment are sampled from a relatively broad distribution.

For the Standard Fisher model, the MLEs for the parameters show only a mild degree of variability between the experiments. 
The difference in initial conditions does not lead to significant differences in the estimates of $D_0$ and $r$ \footnote{A two sample Kolmogorov–Smirnov test could not reject the null hypothesis that the MLEs for the experiments with different initial conditions come from the same distribution, with a $p-$value of $0.1074$ for $D_0$ and $0.5344$ for $r$, respectively.}, while the estimate for $K$ tends to be slightly lower for the experiment with triangular initial conditions as opposed to circular initial conditions \footnote{With a K-S $p-$value of $0.0110$.}.


While the variability of parameter estimates in the Standard Fisher model across experiments is modest, it is much larger in the other models: for example, the standard deviations for the MLEs for $D_0$ across the experiments are 2937.8 for Porous Fisher, 812.9 for Richards, and 320.4 for the Generalised Fisher models, compared to 238.3 for Standard Fisher, although this comparison might not be entirely fair for the Porous Fisher model, since $D_0$ has different units in that model than the others, hence different interpretations. Similar results holds for $r$, while the variability for $K$ is approximately the same across the four models.

\begin{figure}[htbp]
    \centering
    \includegraphics[width=0.55\columnwidth]{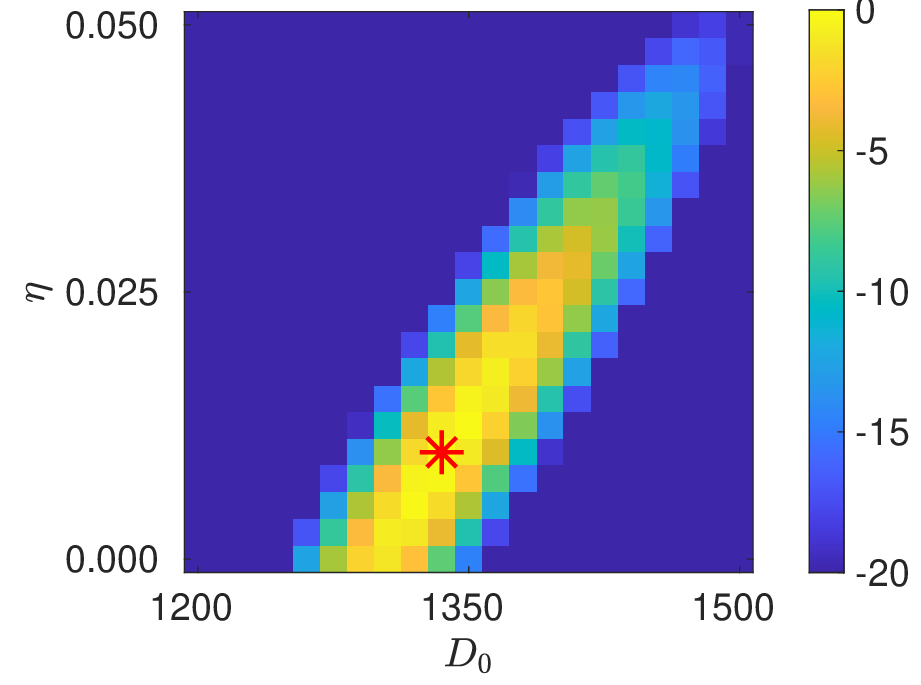}
    \caption[Bivariate likelihood]{Bivariate profile likelihood function for $D_0$ and $r$ in the Porous Fisher model represented as a heat map, computed using the dataset from Experiment 1. The location of the MLE is indicated by the red star. The slant of the function contour indicates that the two variables compensate each other, which reflects a lack of practical identifiability, which could not be detected by examining the univariate profile likelihoods alone.
    }
    \label{fig:bivariate}
\end{figure}

For the Porous Fisher model, the experiments with circular initial conditions (Experiments 1-4) estimate a very small $\eta$, well below $0.2$, suggesting that there is little evidence for nonlinear diffusion in the dynamics, and the estimates for the other parameters $(D_0, r, K)$ are relatively similar to their estimates in the Standard Fisher model. However, the experiments with triangular initial conditions (Experiments 5-8) estimate a much larger $\eta$, along with a higher $D_0$ \footnote{The K-S test shows that the difference in $D_0$ and $\eta$ are significant, while the difference in $r$ and $K$ are not significant.}, suggesting that nonlinear diffusion effects are important. Since the speed of the propagation of the cell colony (which can be identified as the speed of the travelling wave in the model solutions) does not vary much between the experiments, and an increase in $\eta$ corresponds to a decrease in wave speed, the estimated values for $D_0$ and $r$ are much higher for these experiments to compensate for a higher $\eta$. 
This compensation is reflected by the slanted ridge seen in the the bi-variate likelihood function with respect to $D_0$ and $\eta$, shown in Fig.~\ref{fig:bivariate}. 


We see a similar story for the Richards model. Observe that in the third row of Fig.~\ref{fig:conf_interval_compare_all}, there seems to be a clear dependence of the estimated parameter values on the experimental initial conditions. Experiments 1-4 (circular initial conditions) in general give lower values for $D_0, r$, and $\gamma$, and higher values for $K$, compared to Experiments 5-8 (triangular initial conditions). Especially striking is the case of $\gamma$, where Experiments 1-4 give estimates in the range of 0.6-1.7, whereas Experiments 5-8 gave estimates either around eight to nine, or above nine (we use $\gamma=9$ as an upper bound for calculations when searching for the MLE to ensure numerical stability). 
The Richards model is also the only model that has profile likelihoods which do not have a roughly parabolic shape. For Experiments 5 and 6, the profile likelihoods for $\gamma$ are bi-modal (but the peaks are close together, and the valley between them lies above the -1.92 threshold, so the confidence region remains a connected interval). This bi-modality can also be seen in the profile likelihoods for the other parameters, but is much less prominent. For Experiments 7 and 8, the profile likelihood functions for $\gamma$ appear to be monotonically increasing for large $\gamma$ until the imposed upper bound at $\gamma=9$, suggesting that the MLE is either very large or tends to infinity.

A high $\gamma$ (say, above seven) means that the proliferation function $f(C)$ in Eq.~\eqref{eqn:model} is nearly linear for $0 \leq C < K$ except for $C$ near $K$, and it abruptly drops to zero at $C=K$. An even higher $\gamma$ makes little difference to the shape of $f(C)$, and therefore has little effect on the model solution. For this reason the Richards model is known to be locally structurally non-identifiable in the high $\gamma$ regime~\cite{simpson2020PracticalParameterIdentifiabilitya}. This explains the divergence to infinity of the confidence intervals for $\gamma$ for Experiments 7-8. 

It is more difficult to explain why the experiments with  triangular initial conditions suggest a very large $\gamma$, hence nearly linear proliferation, whereas the experiments with circular initial conditions suggest $\gamma$ is relatively close to one, reflecting dynamics closer to logistic growth. There are prior studies making similar observations. In~\cite{jin2016ReproducibilityScratchAssays}, the authors showed that parameter estimates for similar experiments and models depend strongly on the initial cell density, when the shape of the initial cell population is kept constant. In contrast, however, Jin et al.~\cite{jin2018RoleInitialGeometry} found that the shape of the initial cell density profile does not seem to have a major effect on the estimates for the parameters in the Standard Fisher model, which agrees with our observations. 

The variability of the parameter estimates in the Porous Fisher and Richards models, and their dependence on initial conditions, suggests that these two models are mis-specified, since we would not expect such variability if either model is capable of reflecting the true dynamics of the biological process. This mis-specification is likely due to the fact that neither model takes the coupling between tissue mechanics and geometry into account. In an empirical study, Ravasio et al.~\cite{ravasio2015GapGeometryDictates} demonstrated that edge geometry has a significant influence on the rate of wound closure. This effect arises from the exertion of force on the cells by supra-cellular actomyosin cables around the wound edge which, in turn, is regulated by the underlying geometry. 

For the Generalised Fisher model, the computations for the profile likelihoods did not finish within a reasonable time limit for Experiments 5-8, so we only present the parameter estimates for Experiments 1-4 in the last row in Fig.~\ref{fig:conf_interval_compare_all}.  For Experiments 5-8, the optimisation procedures struggled much more to find the correct global optima compared to Experiments 1-4, suggesting that the likelihood landscape is much more rough.


\subsection{Model complexity is limited by data resolution}\label{sec:data_resolution}

\begin{figure}[htbp]
    \centering
    (a)
    \includegraphics[width=0.27\columnwidth]{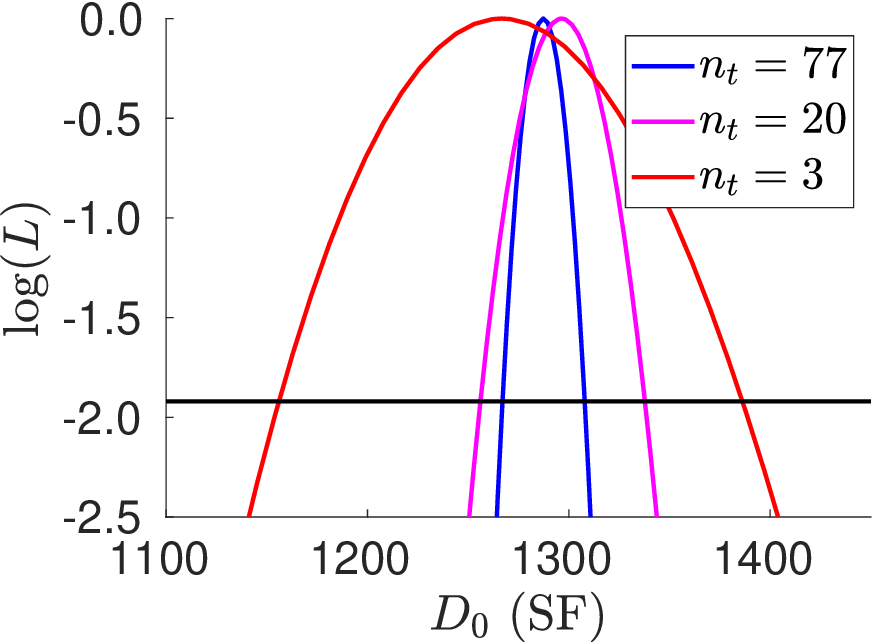}
    (b)
    \includegraphics[width=0.27\columnwidth]{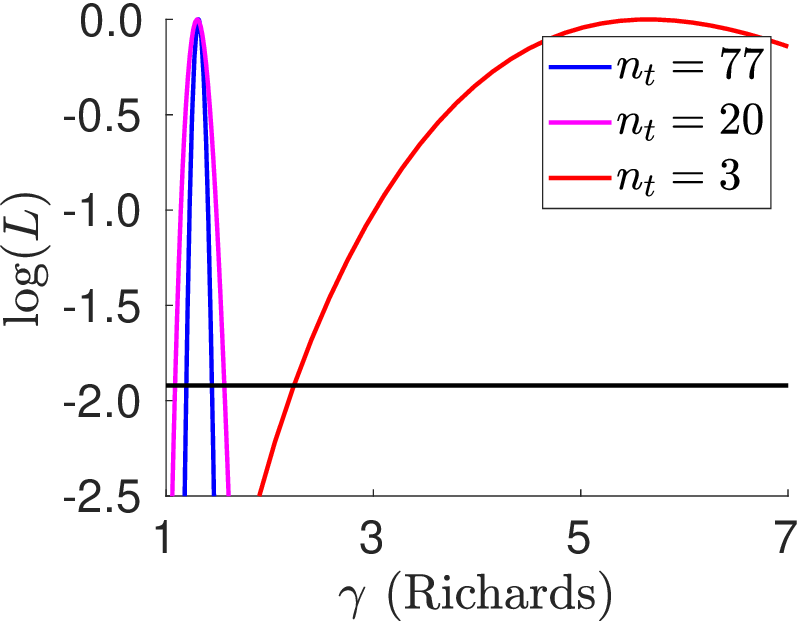}
    (c)
    \includegraphics[width=0.27\columnwidth]{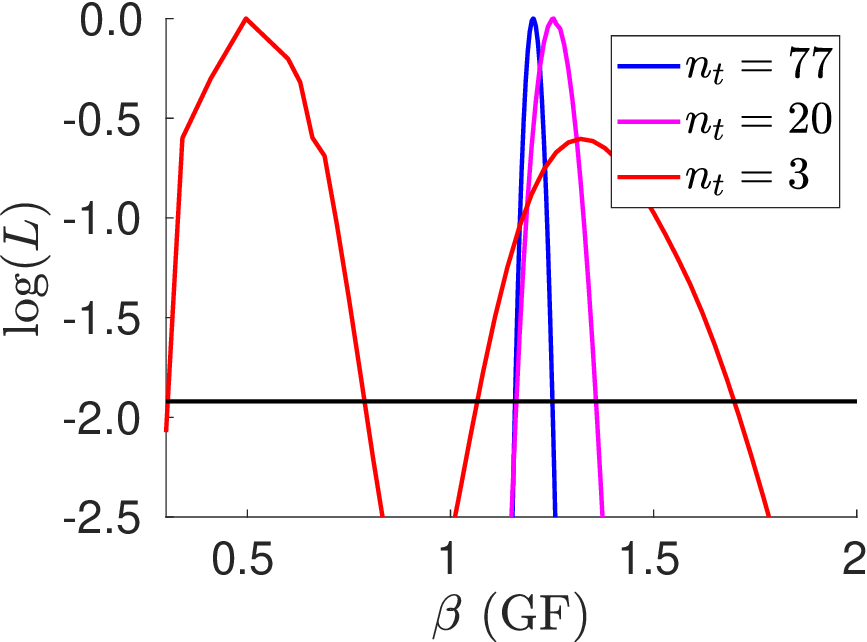}
    \caption[Comparison of profile likelihood curves calculated using progressively down-sampled data]{Comparison of profile likelihood curves calculated using progressively down-sampled data from Experiment 1, for (a) $D_0$ in the Standard Fisher model, (b) $\gamma$ in the Richards model, and (c) $\beta$ in the Generalised Fisher model. $n_t$ denotes the number of time slices in the data we are using. The blue curve corresponds to $n_t=77$, i.e. the profile likelihood calculated with all available data, with a temporal resolution of $\Delta t =20$ min. The pink curve corresponds to $n_t=20$ and $\Delta t=80$ min, and the red curve corresponds to $n_t=3$ and $\Delta t=12$ h. The black horizontal line at $-1.92$ is the threshold for the confidence interval.
    Note that in (b), the red curve remains above the threshold as $\gamma \to \infty$, and we chose to truncate the plot at $\gamma=7$ for ease of visualization.}
    \label{fig:lowdata_pl_compare}
\end{figure}

In order to investigate the relationship between parameter identifiability and data resolution, we repeat the profile likelihood calculation on the reduced dataset from Experiment 1, but progressively down-sample the data to reduce temporal resolution by  utilising only an equally-spaced subset of the 77 time slice in our data. We observe three different ways in which the profile likelihood curves change qualitatively as data resolution decreases. We present the profile likelihood for one parameter to illustrate each case, and leave the rest of the parameters to Supplementary Material Section~\ref{apx:lowdata}. In the first case, the peak of the profile likelihood curve (i.e.~the MLE) remains mostly in the same location, while the curve itself broadens, but nonetheless remains unimodal. This is illustrated by $D_0$ in the Standard Fisher model, shown in Fig.~\ref{fig:lowdata_pl_compare}(a). In this case, despite increases in the uncertainty in parameter estimates, the parameter remains practically identifiable even at the lowest data resolution we considered. This is the case for all parameters in the Standard Fisher model, as well as $K$ in all models, and $D_0$, and $r$ except in the Generalised Fisher model. 

In the second case, as illustrated by $\gamma$, a shape parameter in the Richards model in Fig.~\ref{fig:lowdata_pl_compare}(b), the location of the MLE dramatically changes, and the shape of the profile likelihood curve changes in such a way that the confidence interval becomes infinite, making $\gamma$ non-identifiable in the case where data resolution is low. Interestingly, $\gamma$ is the only parameter that exhibits this behaviour.

The final case is shown in Fig.~\ref{fig:lowdata_pl_compare}(c), where the profile likelihood curve for $\beta$ in the Generalised Fisher model becomes bimodal when data resolution is sufficiently low. $D_0, r$ and $\alpha$ in the Generalised Fisher model also exhibit this behaviour. 


\begin{figure}[htbp]
    \centering
    (a)
    \includegraphics[width=0.27\columnwidth]{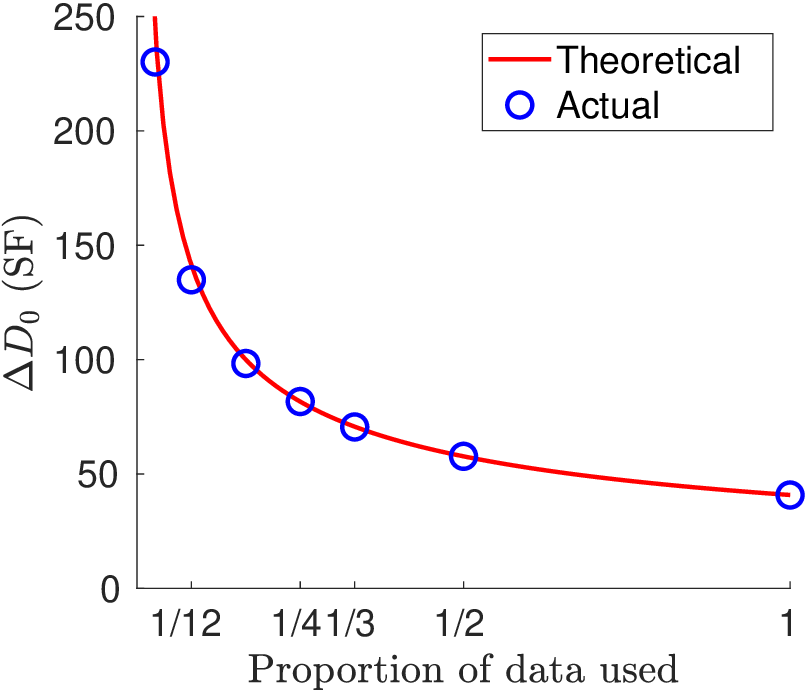}
    (b)
    \includegraphics[width=0.27\columnwidth]{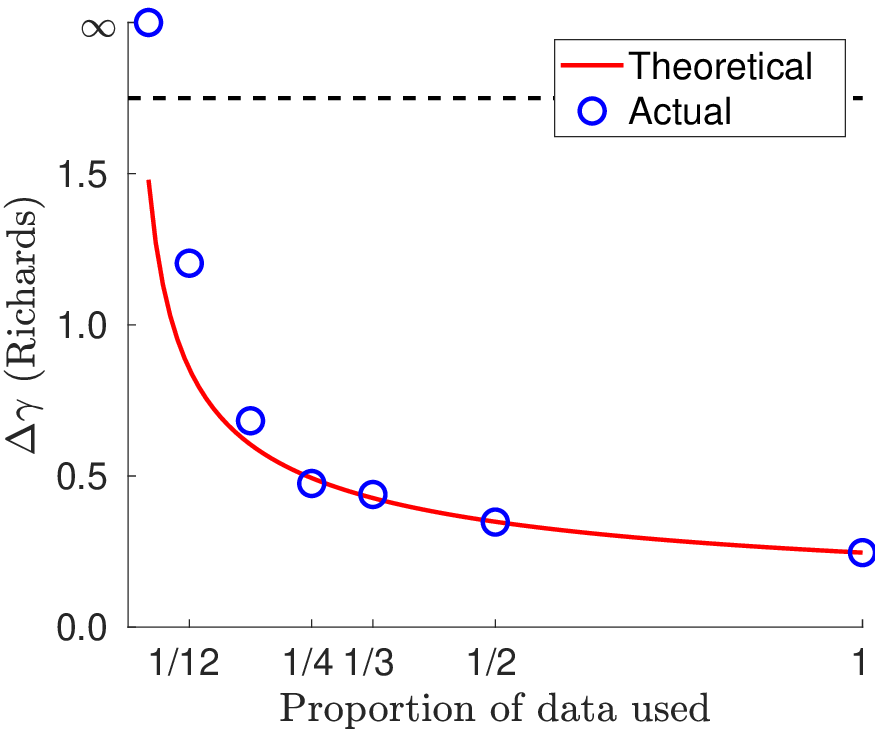}
    (c)
    \includegraphics[width=0.27\columnwidth]{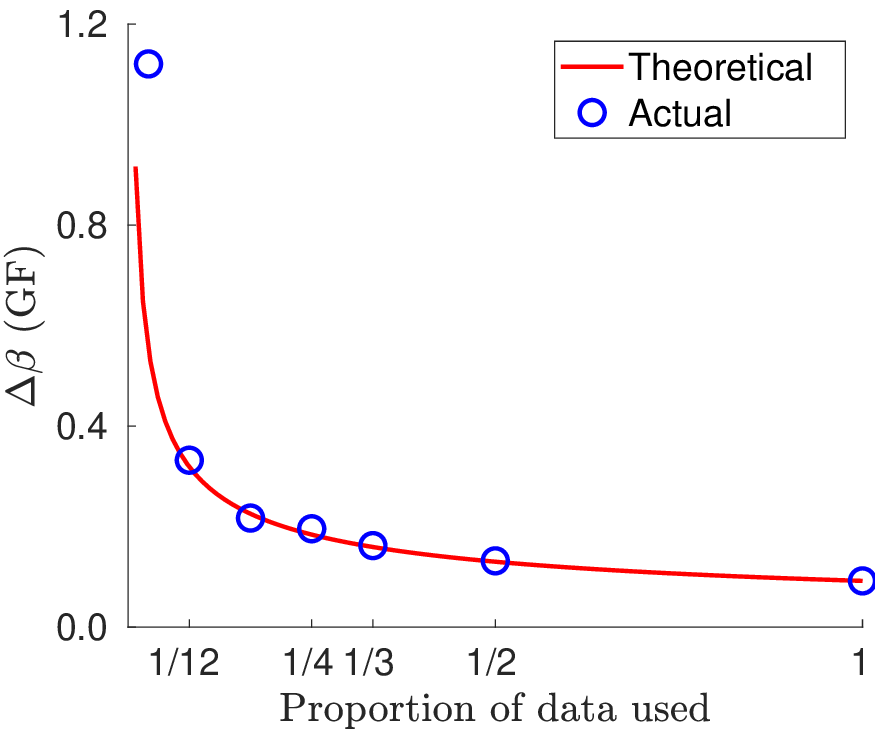}
    \caption[The widths of the confidence intervals as a function of the amount of available data]{The widths of the confidence intervals of three selected parameters plotted against the proportion of data used, for (a) $D_0$ in the Standard Fisher model, (b) $\gamma$ in the Richards model, and (c) $\beta$ in the Generalised Fisher model. 
    Here, using a proportion of $1/m$ of data means that only one in every $m$ images were used, so $n_t=\ceil{77/m}$.
    The blue circles represent the widths of the confidence intervals calculated by finding the intersection between the profile likelihood curves and the threshold -1.92, while the red line is proportional to $1/\sqrt{n_t}$ and normalized so that it goes through the point representing the case where all data were used.}
    \label{fig:conf_interval_width}
\end{figure}

Theoretically, we expect the width of the confidence interval for a parameter $\theta$, denoted $\Delta \theta$, to be proportional to $1/\sqrt{N}$, where $N$ is the number of data points. We compare the theoretical and calculated $\Delta \theta$'s in Fig.~\ref{fig:conf_interval_width}. Observe that for $D_0$ in the Standard Fisher model, the calculated $\Delta D_0$ remain close to the theoretical prediction as $N \to 0$, while for the other two parameters, $\Delta \gamma$ and $\Delta \beta$ deviate significantly from the theoretical prediction as $N \to 0$.

These observations show that parameter identifiability is limited by data resolution. There is a model-dependent minimum amount of data required for the model to be practically identifiable, and this minimum increases as model complexity increases. Therefore, model selection should be tied to the amount of data available, and as a criterion for choosing a model, we should only use models for which we have sufficient data to make it identifiable.


\section{Discussion and future work}\label{sec:discussion}

In this paper we have carried out an in-depth study of practical parameter identifiability for four related PDE models of cell invasion using the profile likelihood method. We have shown that, given sufficient data, the univariate profile likelihoods of all model parameters are unimodal, with the corresponding confidence intervals relatively narrow (with the exception of the Richards model for datasets with triangular initial conditions), suggesting that the parameters are practically identifiable. Moreover, results obtained using synthetic data (Supplementary Material Section~\ref{apx:synthetic}) demonstrate that the profile likelihood method is capable of recovering the true parameter values in the absence of model mis-specification.

We explored the effects of several aspects of data quantity and quality on parameter identifiability. First, by comparing the profile likelihoods obtained using the full density profile to the ones obtained using the radially-averaged density profile, we found that the parameter estimates obtained for the Standard Fisher model are approximately the same for the two cases, but differ considerably for the more complicated models.
This result suggests that simpler models are more robust against subtle changes in data representation. For such models, it is therefore safe to perform inference using the radially averaged dataset to reduce computational costs, whereas the full dataset should be used for parameter inference for the more complicated models which are more sensitive to changes introduced by data processing.

We have shown that for any parameter, the confidence intervals obtained for a single experiment are narrower relative to the spread of the estimated values for the same parameter across experiments. This result suggests that a nonlinear mixed effects framework may be appropriate. In this view, each experimental replicate corresponds to a parameter set sampled from a specified distribution. However, significantly more experiment replicates will be required to determine a suitable form for these random effects. 

For the more complicated models, we observed a dependency of the parameter estimates on the experimental initial conditions. This is at odds with our assumption that the only mechanisms driving tissue expansion are cell motility and proliferation, and that the only factor modulating these mechanisms is cell density. A likely explanation for this dependency is that mechanical properties of the cell tissue may play a significant role in cell invasion.  As such, an interesting avenue for future research entails including mechanical effects at the leading edge in the models.

Another model assumption worth questioning is the independence of observation noise at every point. Reviewing the data, we observe localised bursts of proliferation activity, seemingly at random. This suggests that stochastic models might be more appropriate. However, inference for stochastic models introduces computational challenges, making them unwieldy for many situations. A potentially useful approach is to retain the deterministic model, but assume that the observation noise at points close to each other is correlated, with the magnitude of the noise possibly time-dependent. This approach entails prescribing a correlation kernel with additional parameters to be fitted, which increases complexity, and therefore increases computational cost and likely decreases identifiability, but with the benefit of being more realistic. 
This approach is discussed in \cite{gill2022RobustSimulationDesign} in the context of generalised linear models. Another approach that could be taken is to avoid making assumptions on the noise by replacing the likelihood function (which must be formulated with a noise model in mind) by a generalised profile likelihood function, which relies less on the noise model, but instead must be calibrated by bootstrapping. The details of this approach are discussed in \cite{warne2023GeneralisedLikelihoodProfiles}.

Our findings reveal the pivotal impact of data resolution on parameter identifiability. As the amount of available data decreases, the confidence intervals for all parameters widen. However, certain parameters remain practically identifiable even at low data resolution, while others become non-identifiable if data resolution is sufficiently low. This observation indicates that each model has its own minimum data requirement for practical identifiability, which tends to be higher for more complicated models.

Finally, we return to the problem of model selection. For most datasets, AIC and BIC select the same model as each other, which is usually the Richards or the Generalised Fisher models (see Supplementary Material Section~\ref{apx:mle_tables} for details). However, as we have seen, these models require more data to be identifiable, and they are also more sensitive to subtle changes in the data introduced by data processing (e.g.~full density profile versus radial averaging). Therefore, we argue that model selection methods should also take the amount of available data into account. For the problem considered in this work, more complicated models can be considered if there are sufficient data to render them practically identifiable, whereas the Standard Fisher model should be favoured if the data resolution is low, or if we have only radially averaged data.

In the future, we will extend our investigation to optimise experimental design with the aim to generate the most useful data for identifying model parameters. 
It would also be worthwhile to extend the scope of the investigation to stochastic models, which can be more suitable for systems with significant process noise. Finally, for cell invasion, it would be interesting to see if the addition of cell cycle data, such as those obtained via the FUCCI reporter, can enhance parameter identifiability, or if the additional complexity introduced by including the cell cycle in the model hinders parameter identifiability instead.


\appendix

\section*{Statements}

\textbf{Data and code availability:} The data, including both the images and the results of image analysis, are provided on Zenodo at \url{https://doi.org/10.5281/zenodo.8377953}. The codes for performing the analysis are provided on Github at \url{https://github.com/liuyue002/woundhealing}.

\bigskip\noindent
\textbf{Contribution:} Conceptualisation - REB. Experiments - KS. Data curation - YL, KS. Investigation - YL, KS. Analysis, visualisation - YL. Writing - YL, REB, PKM. Supervision - DJC, REB, PKM.

\bigskip\noindent
\textbf{Competing interests:} We declare no competing interests.

\bigskip\noindent
\textbf{Funding:} YL is supported by the Natural Sciences and Engineering Research Council of Canada (NSERC) through the Postgraduate Scholarships – Doctoral program, reference number PGSD3-535584-2019, as well as by the Canadian Centennial Scholarship Fund (CCSF).

\newpage

\begin{center}
    \Huge Supplementary Material\\ Parameter identifiability and model selection for PDE models of cell invasion \\
    \large Yue Liu$^{1*}$, Kevin Suh$^{2}$, Philip K. Maini$^{1}$, Daniel J. Cohen$^{2,3}$, Ruth E. Baker$^{1}$ \\
$^{1}$Mathematical Institute, University of Oxford\\
$^{2}$Department of Chemical and Biological Engineering, Princeton University\\
$^{3}$Department of Mechanical and Aerospace Engineering, Princeton University\\
$^*$ Corresponding author: yue.liu@maths.ox.ac.uk
\end{center}

\newpage

\section{Numerical methods}

This section provides details of the numerical methods for model simulation and calculation of the profile likelihoods.


\subsection{Numerical solutions of the PDE models}\label{apx:fdm}

We use a finite difference method to simulate the general form of the model, given in Eq.~\eqref{eqn:model}. For simulations in two spatial dimensions, the size of the domain, corresponding to the size of the image, is $L_x=L_y=4380$ \mum. This domain is discretized into $n_x=150$ by $n_y=150$ squares, each with side length $\Delta x = \Delta y = 29.2$ \mum. We used $\Delta t =1/30$ h.

Let $C_{i,j,k}$ denote $C(x_i,y_j,t_k)$, where $x_i=(i-1) \Delta x$, $y_j= (j-1) \Delta y$, $t_k=(k-1) \Delta t$ are the mesh points. The scheme we used follows =~\cite{warne2019UsingExperimentalData}, and can be written as follows:
\begin{align}
\pd{}{x}\left[ D(C_{i,j,k})\pd{C_{i,j,k}}{x} \right] &\approx \frac{1}{\Delta x} \left[D(C_{i+1/2,j,k}) \frac{C_{i+1,j,k}-C_{i,j,k}}{\Delta x} - D(C_{i-1/2,j,k}) \frac{C_{i,j,k}-C_{i-1,j,k}}{\Delta x} \right] \notag\\
&= \frac{1}{(\Delta x)^2} \bigg[D(C_{i+1/2,j,k}) C_{i+1,j,k} - (D(C_{i+1/2,j,k})+D(C_{i-1/2,j,k})) C_{i,j,k} \notag\\
&\quad + D(C_{i-1/2,j,k}) C_{i-1,j,k} \bigg] \tag{SM.1}\label{eqn:space_discr},
\end{align}
where \[D(C_{i+1/2,j,k}) = \frac{1}{2} \bigg[D(C_{i,j,k})+D(C_{i+1,j,k}) \bigg], \quad 1 \leq i,j \leq n_x=n_y=150, \quad 1 \leq k \leq n_t=77. \]
The discretization in the $y$ direction is completely analogous.
Zero flux boundary conditions are imposed at $x=0, L_x$ and $y=0, L_y$.

We use an implicit-explicit (IMEX) scheme~\cite{ascher1995ImplicitExplicitMethodsTimeDependent,ruuth1995ImplicitexplicitMethodsReactiondiffusion} for time-stepping, where the nonlinear diffusion coefficient, $D(C)$, and the proliferation term, $f(C)$, are treated with the explicit Euler method, and the diffusion term overall is treated with the implicit Crank-Nicolson method, which has second order convergence. The advantage of this scheme is that the explicit treatment of the nonlinear components of the equation allows us to avoid having to solve a nonlinear root-finding problem at every time step, which would be necessitated by a fully implicit scheme. The implicit treatment of the diffusion term improves the stability of the scheme, and~\cite{ascher1995ImplicitExplicitMethodsTimeDependent} showed that this class of schemes has reasonably low relative errors when the diffusion term is not vanishingly small, which is the case in this work.
The IMEX Crank-Nicolson time stepping scheme can be written as
\[
\pd{C_{i,j,k}}{t} \approx \frac{C_{i,j,k+1}-C_{i,j,k}}{\Delta t} = \frac{1}{2}\bigg[\nabla\cdot(D(C_{i,j,k}) \nabla C_{i,j,k+1}) + \nabla\cdot(D(C_{i,j,k}) \nabla C_{i,j,k}) \bigg] + f(C_{i,j,k}). \]

We have verified that the scheme is convergent by successively halving $\Delta x$ or $\Delta t$ and recomputing the model solutions with the default parameter values in Eq.~\eqref{eqn:default_param_val}, and check that the norm of the difference between successive model solutions decreases almost linearly on a log-log plot with respect to $\Delta x$ or $\Delta t$.

To justify that the discretisation we have chosen is sufficiently fine, let $\Cmodel^1$ denote the model solution computed with $\Delta x=29.2$ \mum\ and $\Delta t=1/30$ h, $\Cmodel^2$, $\Cmodel^3$ be the model solution computed with $\Delta x$ or $\Delta t$ halved, respectively. Then the difference between $\Cmodel^1$ and $\Cmodel^2$, averaged over all grid points, is 0.448, while that between $\Cmodel^1$ and $\Cmodel^3$ is 4.224, both much smaller than the averaged magnitude of the model solutions, which is on the order of $10^3$, therefore we conclude that the numerical scheme is suitably accurate.


\subsection{Optmisation procedure for MLE and profile likelihoods}\label{apx:optimization}

To solve the optimisation problems for finding the MLEs and evaluating the profile likelihood functions, we use three algorithms, all implemented in MATLAB: the built-in \textit{fmincon} and \textit{globalsearch}, and Covariance Matrix Adaptation Evolution Strategy (CM-AES)~\cite{hansen1996AdaptingArbitraryNormal}, with the implementation obtained at~\cite{cmaes_website}.

The optimisation procedure is initialized with the following default parameter values:
\begin{equation}
D_0=1300 \text{\ \mum$^2$/h}, \quad r=0.3\ \text{h}^{-1}, \quad K=2600 \text{\ cell/mm}^2, \quad \alpha=\beta=\gamma=1, \quad \eta=0. \tag{SM.2}\label{eqn:default_param_val}
\end{equation}

We impose the following bounds for the parameters to guide the optimisation procedures:
\begin{equation}
\begin{split}
100 \text{\ \mum$^2$/h} <D_0 < 10000 \text{\ \mum$^2$/h},  \quad 0.01 \text{h}^{-1} < r < 1 \text{h}^{-1},  \\ \quad 500 \text{\ cell/mm$^2$} < K < 5000 \text{\ cell/mm$^2$},  \quad 0 < \alpha,\beta,\eta < 3, \quad 0 < \gamma < 9.
\end{split}\tag{SM.3}\label{eqn:param_bounds}
\end{equation}

We use \textit{globalsearch} to find the MLEs, and \textit{fmincon} to evaluate points on the profile likelihood functions. In the case where \textit{fmincon} struggles to find the true maximum, we use CM-AES instead.


\newpage
\section{Profile likelihoods for synthetic datasets}\label{apx:synthetic}

In this section, we present the profile likelihoods for each model for two sets of synthetic data. The main purpose of this exercise is to verify that the profile likelihoods behave as expected under ideal conditions.
The synthetic data are generated by simulating the model, Eq.~\eqref{eqn:model} of the main text, in one spatial dimension, using the parameter values in Eq.~\eqref{eqn:default_param_val}, and perturbing by adding Gaussian noise to the model solution. The ``low noise" dataset uses $\sigma=20$, while the ``high noise" dataset uses $\sigma=400$. 
In comparison, the $\sigma^*$ estimated from real data ranges between $380$ -- $460$, depending on the dataset and the model.

\begin{figure}[H]
    \centering
    \includegraphics[width=0.95\columnwidth]{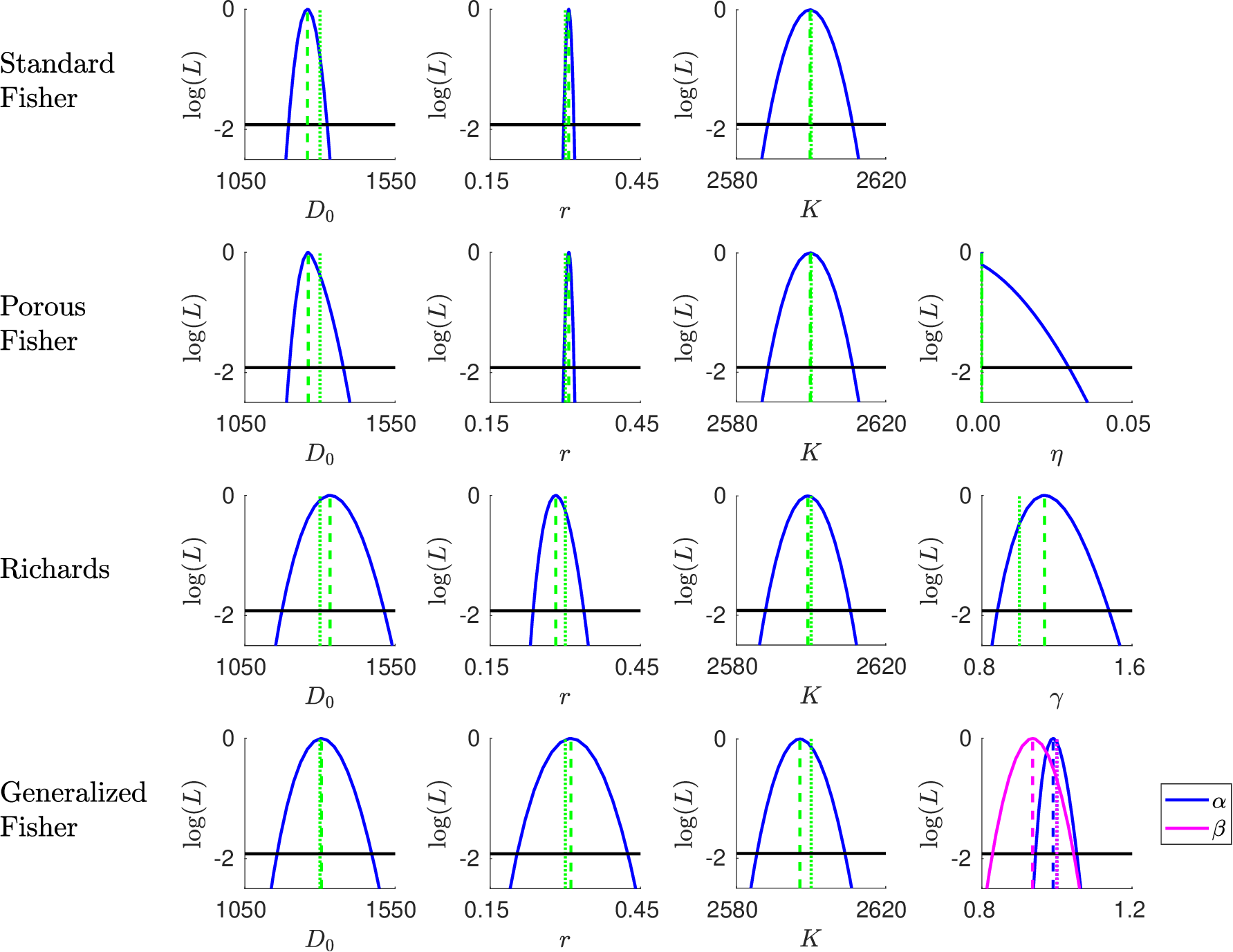}
    \caption[Profile likelihoods for the synthetic dataset]{Profile likelihoods for the four models as described in Eq.~\eqref{eqn:model} and Table~\ref{tab:models}  of the main text, for a synthetic dataset generated with Eq.~\eqref{eqn:model} and parameter values in Eq.~\eqref{eqn:default_param_val}, perturbed as in Eq.~\eqref{eqn:obs_error}  of the main text with $\sigma=400$. The dotted vertical lines mark the location of the true parameter values, while the dashed vertical lines mark the MLE for each parameter. The black horizontal line at $-1.92$ marks the threshold for the 95\% confidence interval. The axis scale for the parameters shared between the models $(D_0, r, K)$ is kept consistent. }
    \label{fig:figtable_syndata}
\end{figure}

The profile likelihoods for the high noise dataset are presented in Fig.~\ref{fig:figtable_syndata}, which shows that all profile likelihood curves are unimodal with a finite confidence interval, and the MLEs are close to the true parameter values.
For the low noise dataset, the profile likelihood curves are very narrow, and centered almost exactly at the true parameter values. These results verify that the profile likelihoods can recover the true parameter values, at least in a highly idealized case, as the theories suggest. 

The profile likelihood curves for the parameters of the Richards and Generalised Fisher models tend to be broader compared to those of the Standard Fisher model, which reflect the greater flexibility of the more complicated models to compensate for a change in one parameter value by shifting the other parameter values.


\newpage
\section{Inference results for all datasets}\label{apx:mle_tables}

In this section, we present the MLE and 95\% confidence intervals calculated for all experimental datasets in table format. Recall that we have cell density data from eight experiments, which we refer to as the full datasets. Experiments 1-4 have circular initial conditions, while Experiments 5-8 have triangular initial conditions.
For Experiments 1-4 we also consider the radially-averaged datasets.
All results are given to four significant figures.


\begin{table}[H]
    \centering
    \resizebox{\columnwidth}{!}{%
    \begin{tabular}{c || c | c | c | c | c | c }
         Model & $D_0$ & $r$ & $K$ & \specialcell{Parameters unique\\ to model} & AIC & BIC\\
         \hline
         \hline
         Standard Fisher & 1287 [1267, 1307] & 0.2707 [0.2683, 0.2731] & 2620 [2614, 2625] & & 162289 & 162310\\ 
         \hline
         Porous Fisher & 1361 [1306, 1419] & 0.2686 [0.2658, 0.2714] & 2622 [2616, 2628] & $\eta: 0.0219 [0.0069, 0.0372]$ & 162283 & 162311\\
         \hline
         Richards & 1467 [1400, 1535] & 0.2272 [0.2146, 0.2410] & 2612 [2606, 2618] & $\gamma: 1.3119 [1.1950, 1.4416]$ & 162258 & 162287\\
         \hline
         Generalised Fisher & 1391 [1321, 1465] & 0.1429 [0.1130, 0.1758] & 2664 [2652, 2678] & \specialcell{$\alpha: 1.1086 [1.0779, 1.1434]$ \\ $\beta: 1.2034 [1.1587, 1.2508]$} & \textbf{162175} & \textbf{162210}\\
    \end{tabular}
    }
    \caption{Experiment 1, radially-averaged dataset}
\end{table}


\begin{table}[H]
    \centering
    \resizebox{\columnwidth}{!}{%
    \begin{tabular}{c || c | c | c | c | c | c }
         Model & $D_0$ & $r$ & $K$ & \specialcell{Parameters unique\\ to model} & AIC & BIC\\
         \hline
         \hline
         Standard Fisher & 1287 [1282, 1293] & 0.2775 [0.2767, 0.2782] & 2621 [2619, 2622] & & 26161393 & 26161430\\ 
         \hline
         Porous Fisher & 1545 [1529, 1564] & 0.2702 [0.2694, 0.2710] & 2628 [2627, 2630] & $\eta: 0.0719 [0.0677, 0.0766]$ & 26160486 & 26160536\\
         \hline
         Richards & 1402 [1385, 1418] & 0.2462 [0.2425, 0.2502] & 2616 [2615, 2618] & $\gamma: 1.1972 [1.1677, 1.2249]$ & 26161172 & 26161221\\
         \hline
         Generalised Fisher & 1423 [1403, 1443] & 0.1013 [0.0945, 0.1085] & 2701 [2698, 2704] & \specialcell{$\alpha: 1.1733 [1.1688, 1.1818]$ \\ $\beta: 1.3548 [1.3437, 1.3663]$} & \textbf{26158023} & \textbf{26158085}\\
    \end{tabular}
    }
    \caption{Experiment 1, full dataset}
\end{table}


\begin{table}[H]
    \centering
    \resizebox{\columnwidth}{!}{%
    \begin{tabular}{c || c | c | c | c | c | c }
         Model & $D_0$ & $r$ & $K$ & \specialcell{Parameters unique\\ to model} & AIC & BIC\\
         \hline
         \hline
         Standard Fisher & 1211 [1192, 1231] & 0.2780 [0.2755, 0.2805] & 2551 [2546, 2556] & & 162216 & 162238\\ 
         \hline
         Porous Fisher & 1650 [1578, 1725] & 0.2673 [0.2645, 0.2701] & 2564 [2558, 2570] & $\eta: 0.1270 [0.1086, 0.1458]$ & 162009 & 162038\\
         \hline
         Richards & 981 [938, 1026] & 0.3675 [0.3474, 0.3891] & 2560 [2554, 2565] & $\gamma: 0.6808 [0.6333, 0.7323]$ & 162115 & 162143\\
         \hline
         Generalised Fisher & 806 [758, 857] & 0.2787 [0.2374, 0.3219] & 2873 [2838, 2913] & \specialcell{$\alpha: 1.0744 [1.0532, 1.0984]$ \\ $\beta: 2.0151 [1.9249, 2.1130]$} & \textbf{160935} & \textbf{160970}\\
    \end{tabular}
    }
    \caption{Experiment 2, radially-averaged dataset}
\end{table}


\begin{table}[H]
    \centering
    \resizebox{\columnwidth}{!}{%
    \begin{tabular}{c || c | c | c | c | c | c }
         Model & $D_0$ & $r$ & $K$ & \specialcell{Parameters unique\\ to model} & AIC & BIC\\
         \hline
         \hline
         Standard Fisher & 1157 [1152, 1163] & 0.2896 [0.2888, 0.2904] & 2550 [2548, 2552] & & 26099801 & 26099838\\ 
         \hline
         Porous Fisher & 1527 [1509, 1545] & 0.2788 [0.2779, 0.2795] & 2561 [2560, 2563] & $\eta: 0.1099 [0.1071, 0.1129]$ & 26097458 & 26097507\\
         \hline
         Richards & 916 [907, 926] & 0.4061 [0.4004, 0.4119] & 2561 [2559, 2562] & $\gamma: 0.6307 [0.6196, 0.6419]$ & 26097356 & 26097406\\
         \hline
         Generalised Fisher & 826 [816, 837] & 0.2383 [0.2283, 0.2484] & 2893 [2881, 2905] & \specialcell{$\alpha: 1.1091 [1.1024, 1.1160]$ \\ $\beta: 2.1168 [2.0860, 2.1485]$} & \textbf{26082314} & \textbf{26082376}\\
    \end{tabular}
    }
    \caption{Experiment 2, full dataset}
\end{table}


\begin{table}[H]
    \centering
    \resizebox{\columnwidth}{!}{%
    \begin{tabular}{c || c | c | c | c | c | c }
         Model & $D_0$ & $r$ & $K$ & \specialcell{Parameters unique\\ to model} & AIC & BIC\\
         \hline
         \hline
         Standard Fisher & 1107 [1083, 1131] & 0.3172 [0.3133, 0.3212] & 2518 [2511, 2525] & & 169220 & 169242\\ 
         \hline
         Porous Fisher & 1228 [1160, 1301] & 0.3136 [0.3093, 0.3179] & 2520 [2513, 2527] & $\eta: 0.0394 [0.0191, 0.0604]$ & \textbf{169207} & \textbf{169236}\\
         \hline
         Richards & 1200 [1117, 1288] & 0.2860 [0.2622, 0.3122] & 2514.7859 [2507, 2522] & $\gamma: 1.1669 [1.0249, 1.3367]$ & 169217 & 169245\\
         \hline
         Generalised Fisher & 1146 [1064, 1237] & 0.2391 [0.1941, 0.2866] & 2534 [2516, 2553] & \specialcell{$\alpha: 1.0501 [1.0231, 1.0811]$ \\ $\beta: 1.1005 [1.0093, 1.1850]$} & 169209 & 169245\\
    \end{tabular}
    }
    \caption{Experiment 3, radially-averaged dataset}
\end{table}


\begin{table}[H]
    \centering
    \resizebox{\columnwidth}{!}{%
    \begin{tabular}{c || c | c | c | c | c | c }
         Model & $D_0$ & $r$ & $K$ & \specialcell{Parameters unique\\ to model} & AIC & BIC\\
         \hline
         \hline
         Standard Fisher & 1118 [1113,1122] & 0.3198 [0.3191,0.3206] & 2521 [2520,2523] & & 26152075 & 26152112\\ 
         \hline
         Porous Fisher & 1300 [1286,1314] & 0.3144 [0.3136,0.3153] & 2524 [2523, 2526] & $\eta: 0.0569 [0.0529,0.0609]$ & 26151214 & 26151263\\
         \hline
         Richards & 1293 [1275,1307] & 0.2633 [0.2597,0.2678] & 2514 [2513,2516] & $\gamma: 1.3493 [1.3131,1.3801]$ & 26151457 & 26151506\\
         \hline
         Generalised Fisher & 1377 [1353,1398] & 0.1999 [0.1919,0.2080] & 2500 [2498,2504] & \specialcell{$\alpha: 1.0506 [1.0439,1.0576]$ \\ $\beta: 0.8671 [0.8443,0.8943]$} & \textbf{26151117} & \textbf{26151179}\\
    \end{tabular}
    }
    \caption{Experiment 3, full dataset}
\end{table}


\begin{table}[H]
    \centering
    \resizebox{\columnwidth}{!}{%
    \begin{tabular}{c || c | c | c | c | c | c }
         Model & $D_0$ & $r$ & $K$ & \specialcell{Parameters unique\\ to model} & AIC & BIC\\
         \hline
         \hline
         Standard Fisher & 1239 [1221,1257] & 0.2849 [0.2825,0.2873] & 2784 [2779,2790] & & 161998 & 162019\\ 
         \hline
         Porous Fisher & 1406 [1356,1458] & 0.2800 [0.2773,0.2827] & 2789 [2784,2795] & $\eta: 0.0499 [0.0363,0.0637]$ & 161945 & 161974\\
         \hline
         Richards & 1466 [1410,1523] & 0.2273 [0.2168,0.2387] & 2775 [2770,2781] & $\gamma: 1.4221 [1.3136,1.5407]$ & 161922 & 161951\\
         \hline
         Generalised Fisher & 1416 [1358,1476] & 0.1464 [0.1216,0.1732] & 2799 [2789,2810] & \specialcell{$\alpha: 1.1009 [1.0761,1.1284]$ \\ $\beta: 1.0869 [1.0457,1.1293]$} & \textbf{161913} & \textbf{161948}\\
    \end{tabular}
    }
    \caption{Experiment 4, radially-averaged dataset}
\end{table}


\begin{table}[H]
    \centering
    \resizebox{\columnwidth}{!}{%
    \begin{tabular}{c || c | c | c | c | c | c }
         Model & $D_0$ & $r$ & $K$ & \specialcell{Parameters unique\\ to model} & AIC & BIC\\
         \hline
         \hline
         Standard Fisher & 1252 [1247,1258] & 0.2865 [0.2858,0.2872] & 2788 [2787,2790] & & 26157597 & 26157634\\ 
         \hline
         Porous Fisher & 1440 [1425,1454] & 0.2809 [0.2801,0.2817] & 2794 [2792,2795] & $\eta: 0.0539 [0.0500,0.0575]$ & 26156816 & 26156866\\
         \hline
         Richards & 1565 [1548,1580] & 0.2108 [0.2084,0.2137] & 2775 [2773,2776] & $\gamma: 1.6398 [1.6013,1.6736]$ & \textbf{26155625} & \textbf{26155675}\\
         \hline
         Generalised Fisher & 1552 [1533,1570] & 0.1028 [0.0965,0.1093] & 2797 [2794,2800] & \specialcell{$\alpha: 1.1489 [1.1398,1.1583]$ \\ $\beta: 1.0701 [1.0572,1.0831]$} & 26155629 & 26155691\\
    \end{tabular}
    }
    \caption{Experiment 4, full dataset}
\end{table}


\begin{table}[H]
    \centering
    \resizebox{\columnwidth}{!}{%
    \begin{tabular}{c || c | c | c | c | c | c }
         Model & $D_0$ & $r$ & $K$ & \specialcell{Parameters unique\\ to model} & AIC & BIC\\
         \hline
         \hline
         Standard Fisher & 1416 [1410,1422] & 0.3085 [0.3077,0.3093] & 2344 [2343,2346] & & 25693899 & 25693936\\ 
         \hline
         Porous Fisher & 4102 [4059,4145] & 0.2739 [0.2731,0.2747] & 2377 [2375,2379] & $\eta: 0.5677 [0.5609,0.5746]$ & 25649571 & 25649620\\
         \hline
         Richards & 2697 [2688,2706] & 0.1364 [0.1359,0.1367] & 2336 [2334,2337] & $\gamma: 7.8055 [7.7527,7.9284]$ & \textbf{25589713} & \textbf{25589762}\\
         \hline
         Generalised Fisher & - & - & - & \specialcell{$\alpha: -$ \\ $\beta: -$} & - & -\\
    \end{tabular}
    }
    \caption{Experiment 5, full dataset}
\end{table}


\begin{table}[H]
    \centering
    \resizebox{\columnwidth}{!}{%
    \begin{tabular}{c || c | c | c | c | c | c }
         Model & $D_0$ & $r$ & $K$ & \specialcell{Parameters unique\\ to model} & AIC & BIC\\
         \hline
         \hline
         Standard Fisher & 1161 [1156,1166] & 0.3244 [0.3235,0.3253] & 2307 [2305,2308] & & 25614660 & 25614697\\ 
         \hline
         Porous Fisher & 2713 [2685,2743] & 0.2931 [0.2922,0.2940] & 2333 [2331,2334] & $\eta: 0.4097 [0.4041,0.4157]$ & 25587515 & 25587565\\
         \hline
         Richards & 2290 [2282,2299] & 0.1413 [0.1409,0.1416] & 2288 [2286,2289] & $\gamma: 8.1042 [8.0335,8.1723]$ & \textbf{25534767} & \textbf{25534816}\\
         \hline
         Generalised Fisher & - & - & - & \specialcell{$\alpha: -$ \\ $\beta: -$} & - & -\\
    \end{tabular}
    }
    \caption{Experiment 6, full dataset}
\end{table}


\begin{table}[H]
    \centering
    \resizebox{\columnwidth}{!}{%
    \begin{tabular}{c || c | c | c | c | c | c }
         Model & $D_0$ & $r$ & $K$ & \specialcell{Parameters unique\\ to model} & AIC & BIC\\
         \hline
         \hline
         Standard Fisher & 1845 [1837,1853] & 0.2260 [0.2254,0.2266] & 2419 [2417,2421] & & 25658251 & 25658288\\ 
         \hline
         Porous Fisher & 10061 [9894,10232] & 0.1637 [0.1628,0.1646] & 2627 [2622,2631] & $\eta: 1.0443 [1.0313,1.0573]$ & 25590211 & 25590260\\
         \hline
         Richards & 3180 [3170,3190] & 0.1057 [0.1055,0.1059] & 2353 [2352,2354] & $\gamma: \infty *$ & \textbf{25521509} & \textbf{25521559}\\
         \hline
         Generalised Fisher & - & - & - & \specialcell{$\alpha: -$ \\ $\beta: -$} & - & -\\
    \end{tabular}
    }
    \caption{Experiment 7, full dataset. Note that for $\gamma$ in the Richards model, the profile likelihood seems to be monotonically increasing up to the upper bound of $\gamma=9$ which we have imposed for numerical stability, therefore the true MLE is likely to be very large or infinite.}
\end{table}


\begin{table}[H]
    \centering
    \resizebox{\columnwidth}{!}{%
    \begin{tabular}{c || c | c | c | c | c | c }
         Model & $D_0$ & $r$ & $K$ & \specialcell{Parameters unique\\ to model} & AIC & BIC\\
         \hline
         \hline
         Standard Fisher & 1448 [1442,1454] & 0.2669 [0.2662,0.2676] & 2294 [2292,2296] & & 25536527 & 25536564\\ 
         \hline
         Porous Fisher & 3504 [3467,3543] & 0.2374 [0.2367,0.2382] & 2337 [2335,2339] & $\eta: 0.4475 [0.4414,0.4539]$ & 25508165 & 25508214\\
         \hline
         Richards & 2666 [2658,2675] & 0.1199 [0.1196,0.1201] & 2241 [2239,2242] & $\gamma: \infty *$ & \textbf{25444985} & \textbf{25445034}\\
         \hline
         Generalised Fisher & - & - & - & \specialcell{$\alpha: -$ \\ $\beta: -$} & - & -\\
    \end{tabular}
    }
    \caption{Experiment 8, full dataset. Similar observations for $\gamma$ as in Experiment 7.}
\end{table}


We also present the profile likelihoods for Experiments 2-8 (those for Experiment 1 are presented in Fig.~\ref{fig:figtable_xy1_2d} and Fig.~\ref{fig:figtable_xy1_1d} of the main text).


\begin{figure}[H]
    \centering
    \includegraphics[width=0.95\columnwidth]{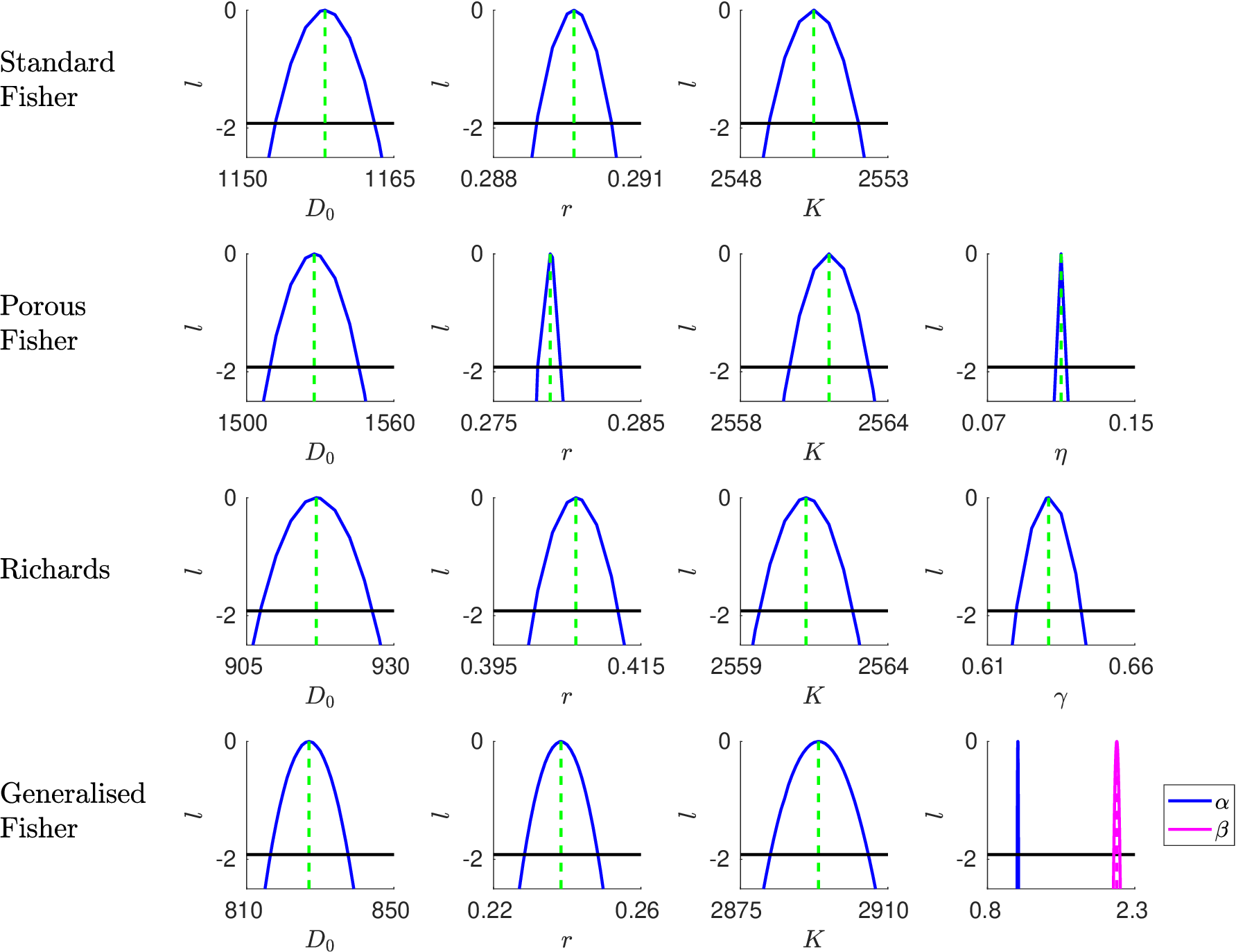}
    \caption{Profile likelihoods for Experiment 2, full dataset. }
\end{figure}


\begin{figure}[H]
    \centering
    \includegraphics[width=0.95\columnwidth]{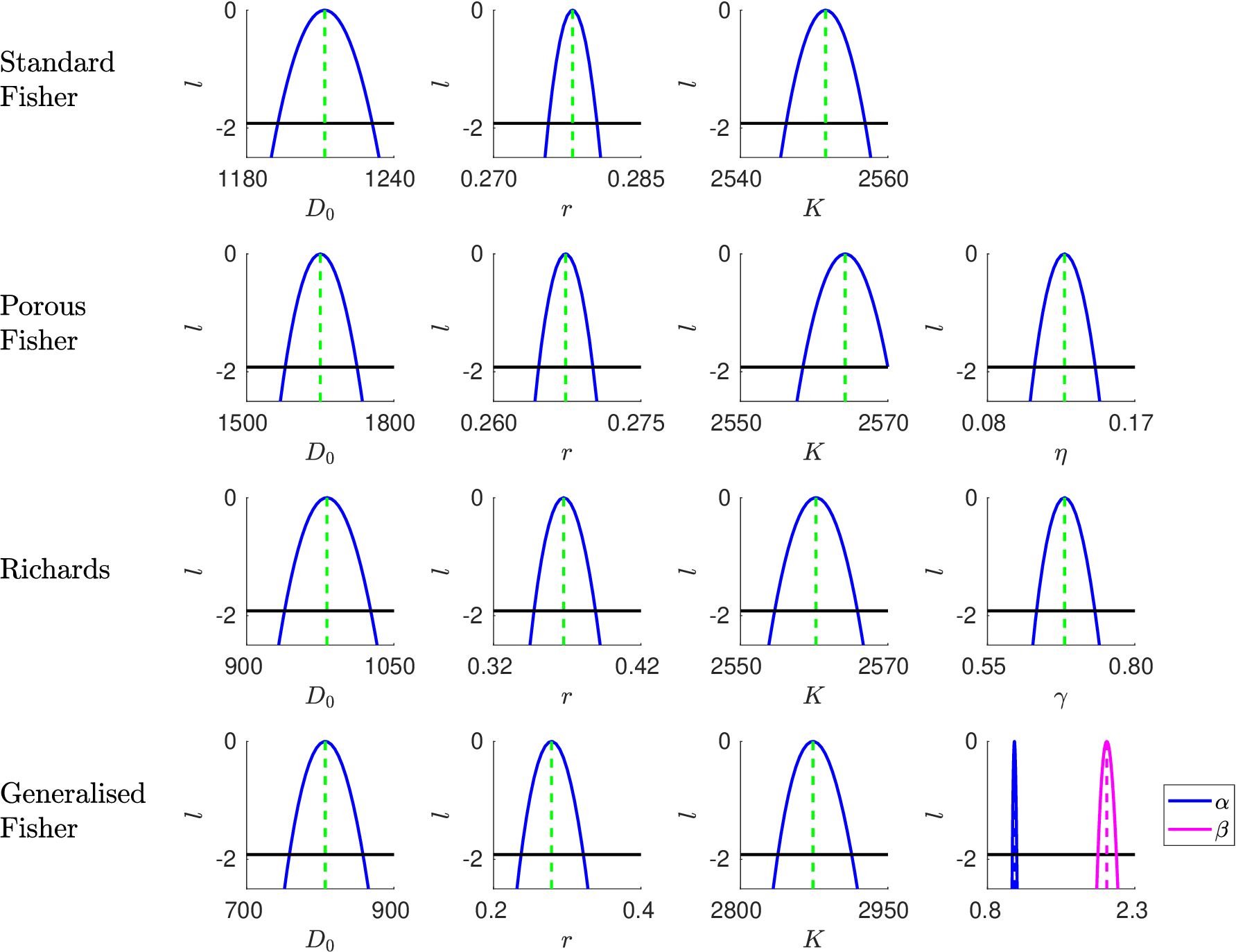}
    \caption{Profile likelihoods for Experiment 2, radially averaged dataset. }
\end{figure}


\begin{figure}[H]
    \centering
    \includegraphics[width=0.95\columnwidth]{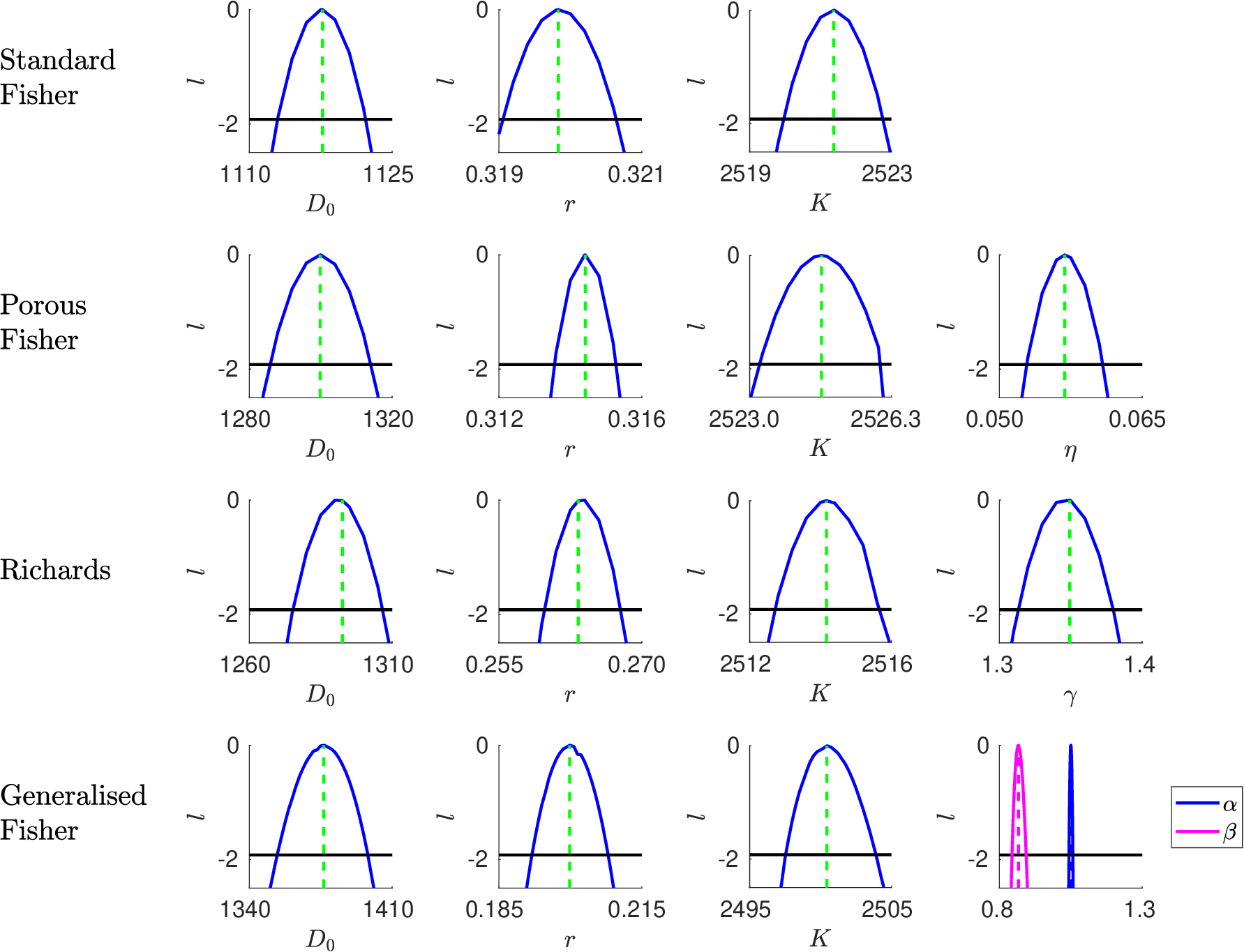}
    \caption{Profile likelihoods for Experiment 3, full dataset. }
\end{figure}


\begin{figure}[H]
    \centering
    \includegraphics[width=0.95\columnwidth]{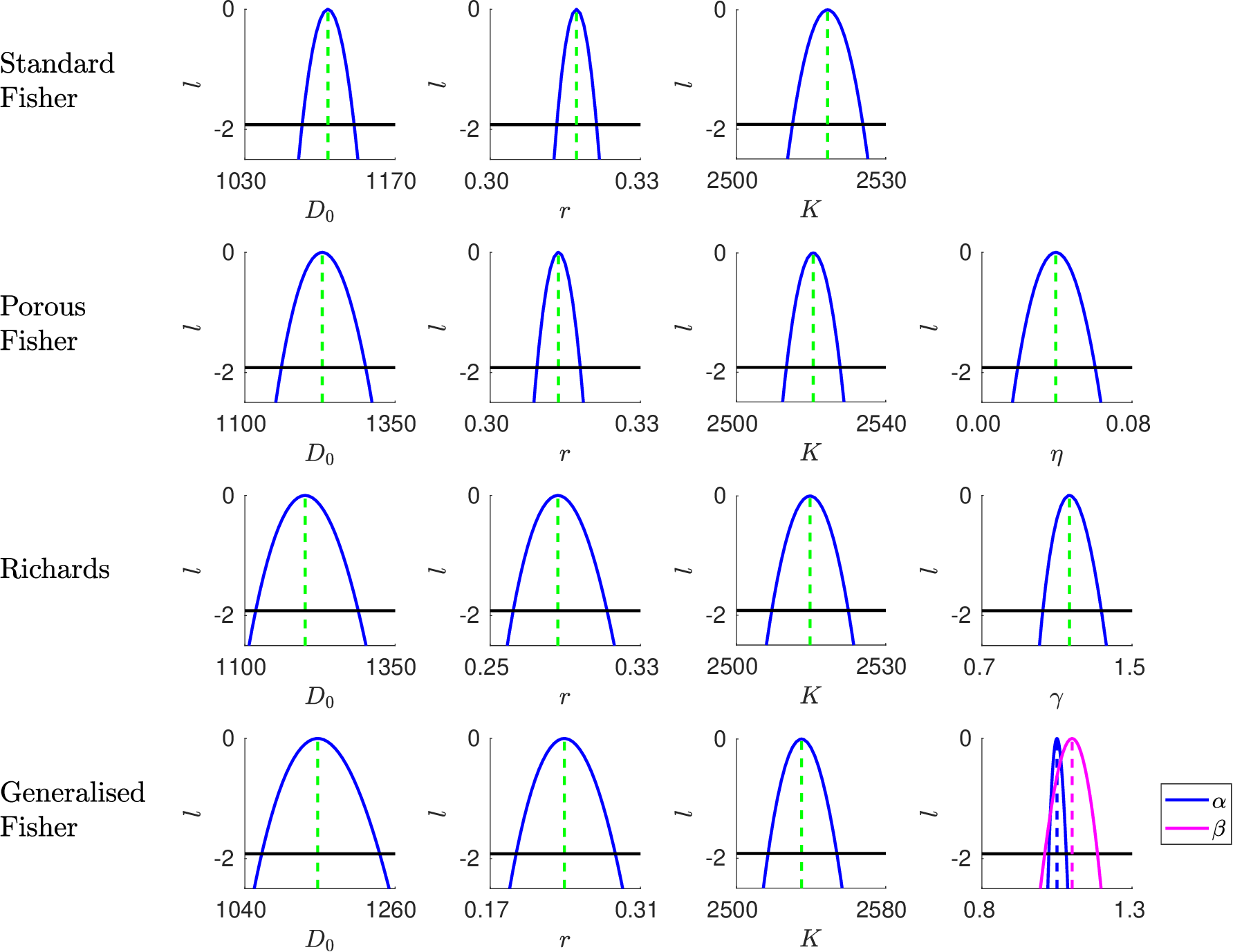}
    \caption{Profile likelihoods for Experiment 3, radially averaged dataset. }
\end{figure}


\begin{figure}[H]
    \centering
    \includegraphics[width=0.95\columnwidth]{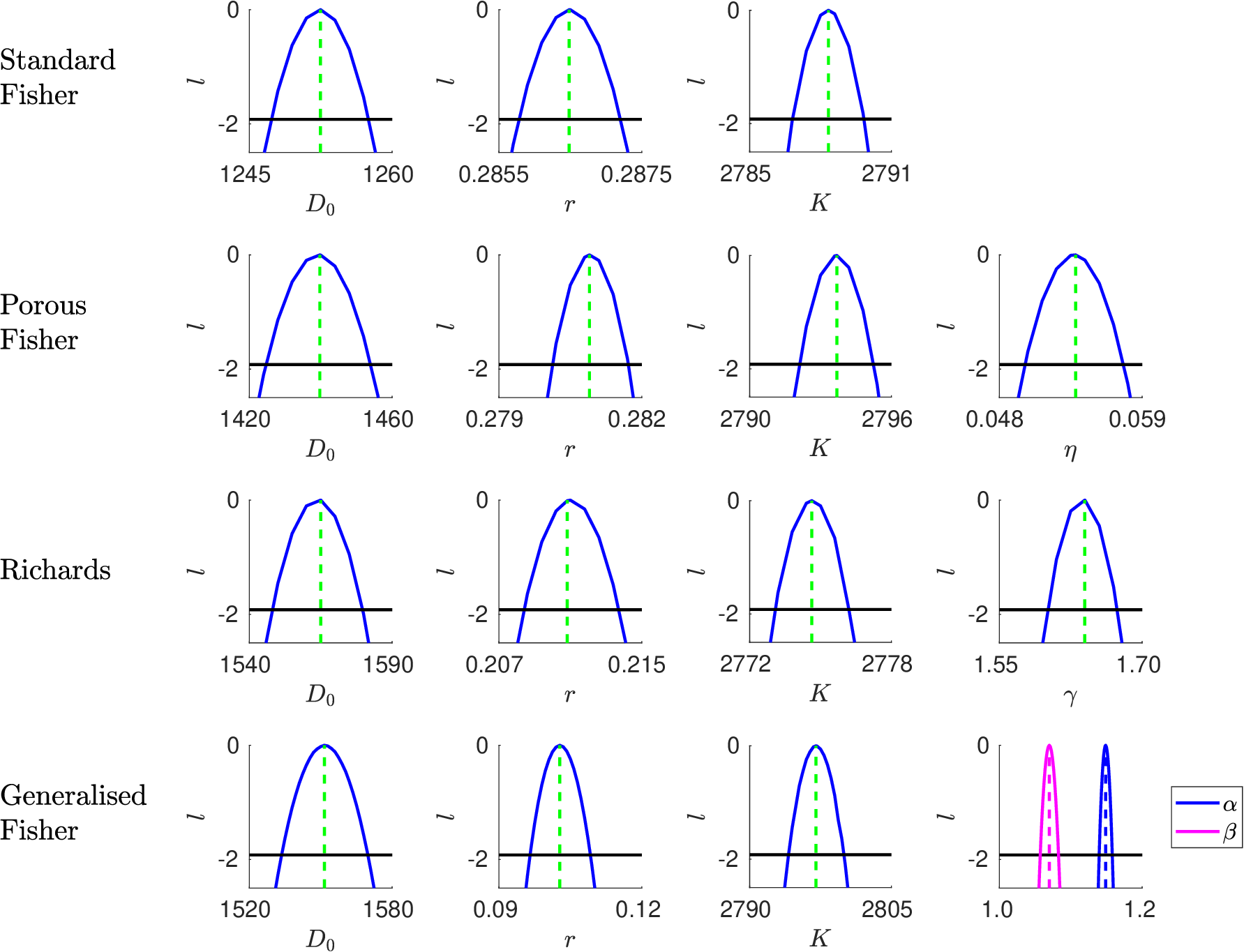}
    \caption{Profile likelihoods for Experiment 4, full dataset. }
\end{figure}


\begin{figure}[H]
    \centering
    \includegraphics[width=0.95\columnwidth]{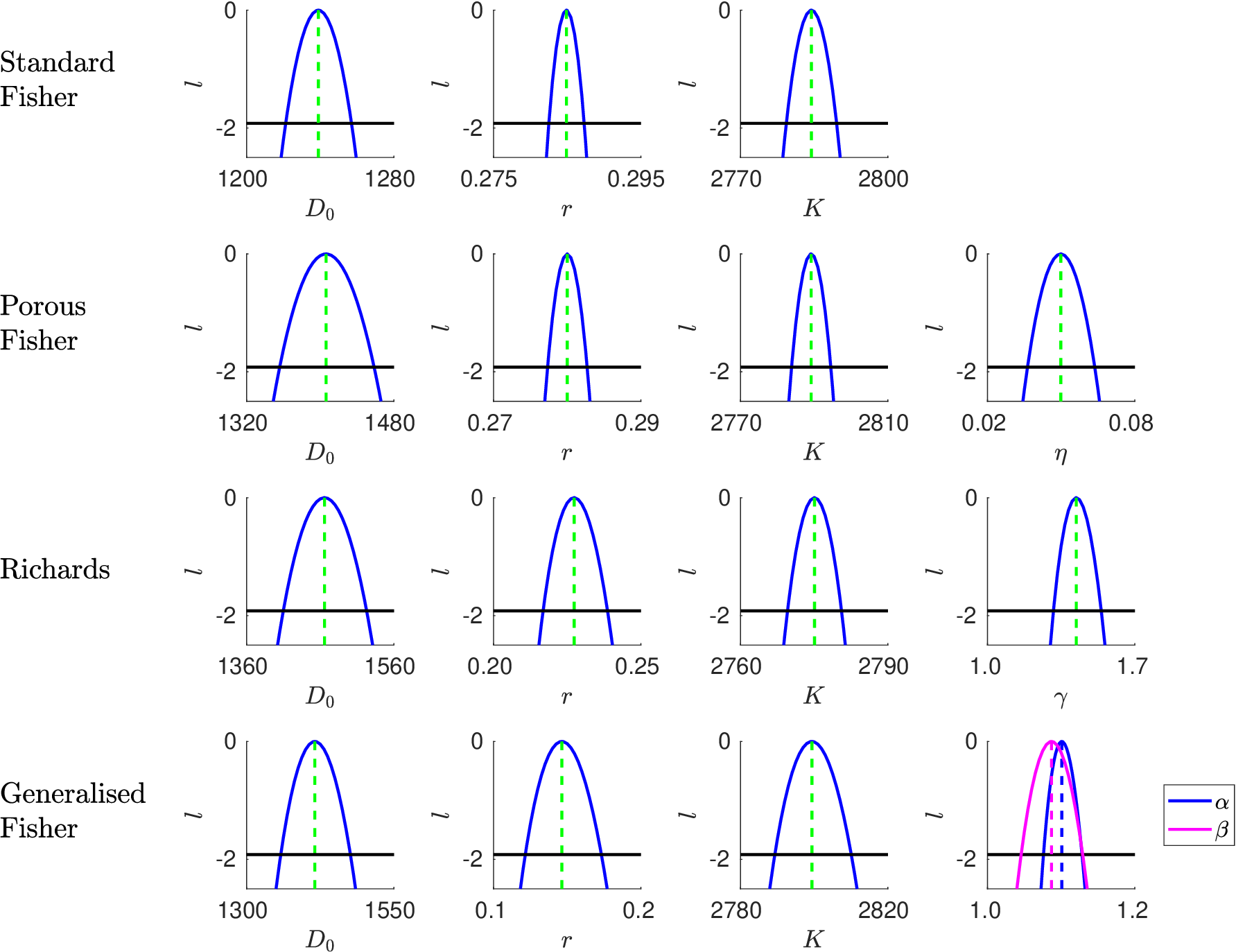}
    \caption{Profile likelihoods for Experiment 4, radially averaged dataset. }
\end{figure}


\begin{figure}[H]
    \centering
    \includegraphics[width=0.95\columnwidth]{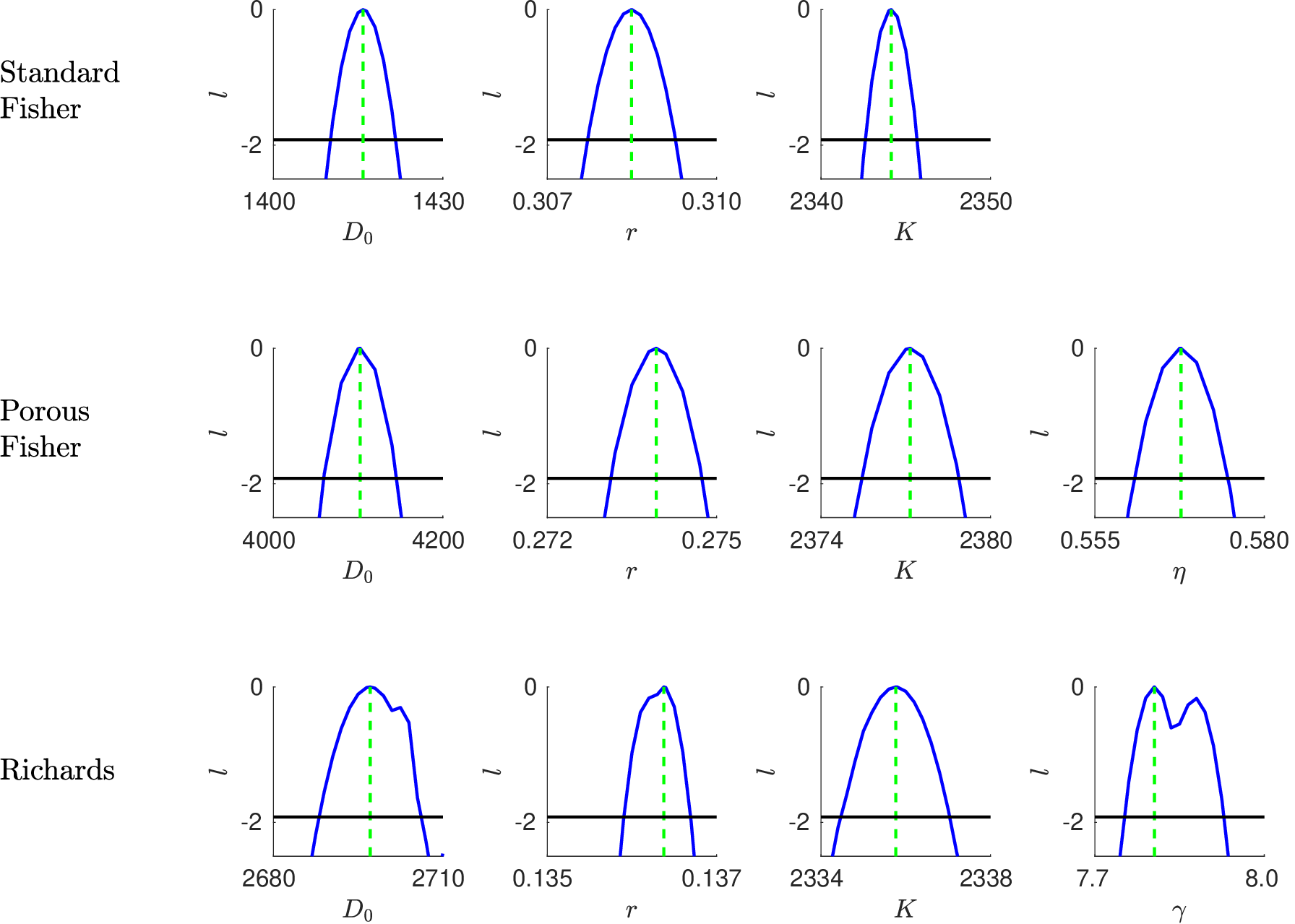}
    \caption{Profile likelihoods for Experiment 5, full dataset. }
\end{figure}


\begin{figure}[H]
    \centering
    \includegraphics[width=0.95\columnwidth]{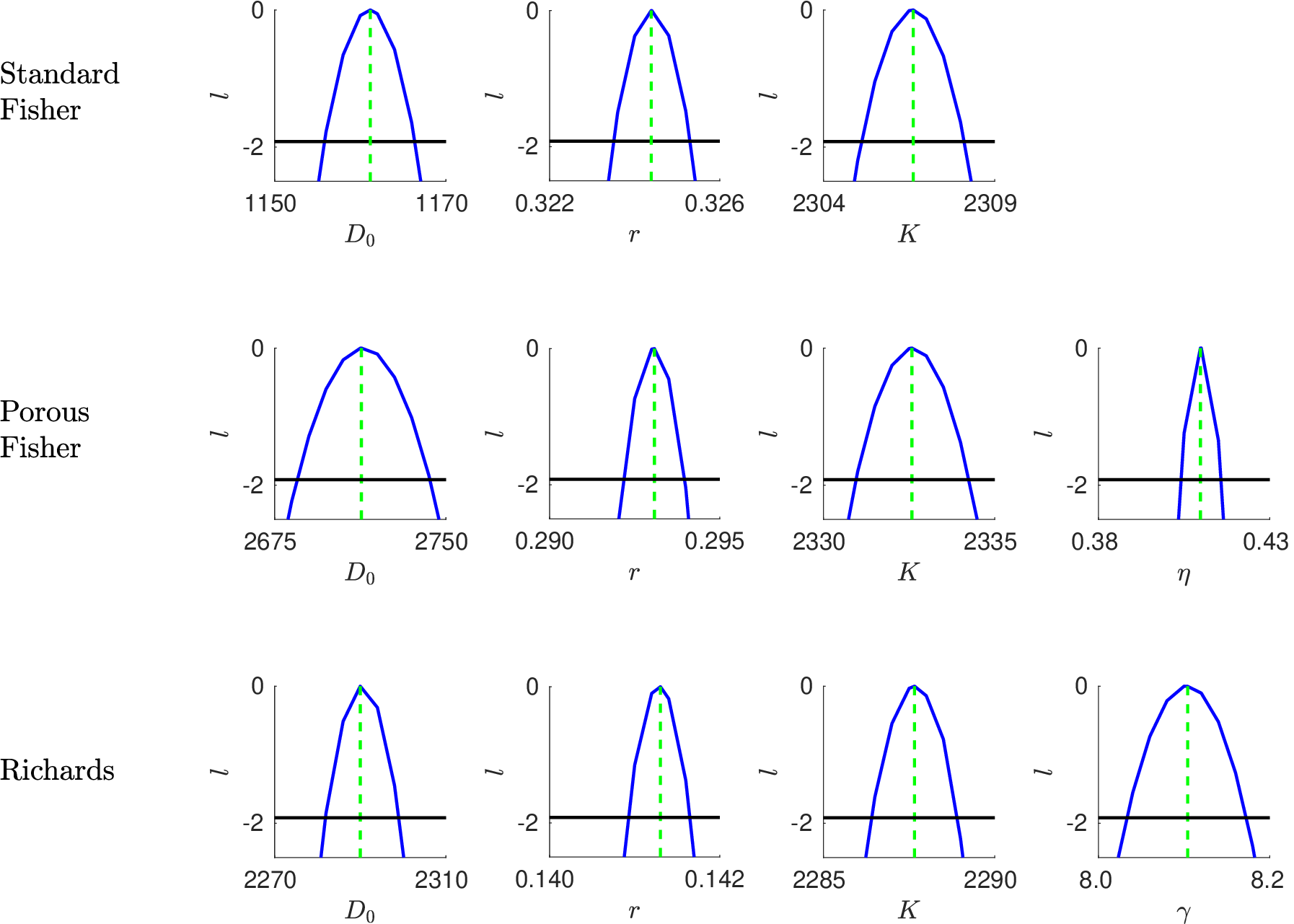}
    \caption{Profile likelihoods for Experiment 6, full dataset. }
\end{figure}


\begin{figure}[H]
    \centering
    \includegraphics[width=0.95\columnwidth]{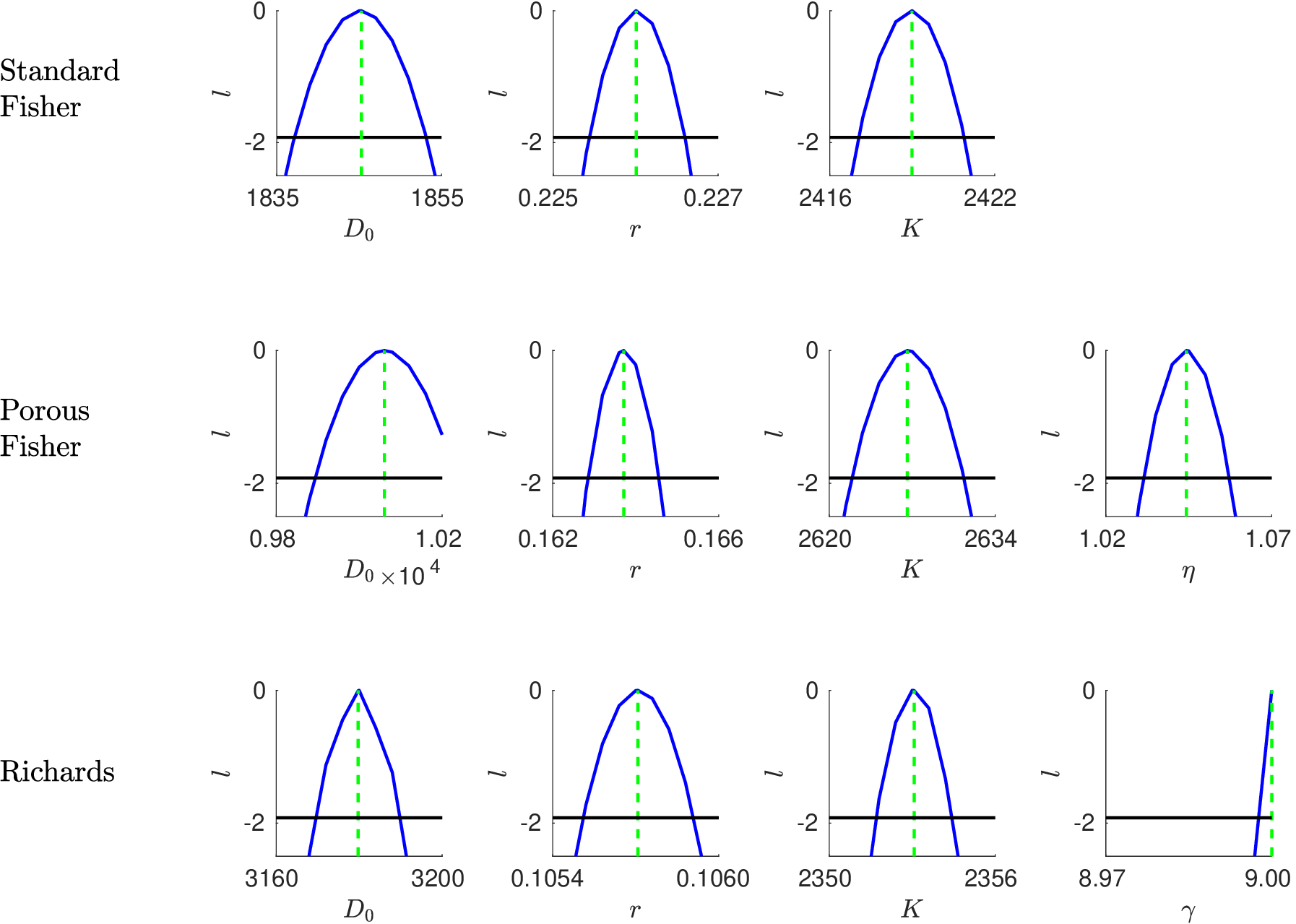}
    \caption{Profile likelihoods for Experiment 7, full dataset. }
\end{figure}


\begin{figure}[H]
    \centering
    \includegraphics[width=0.95\columnwidth]{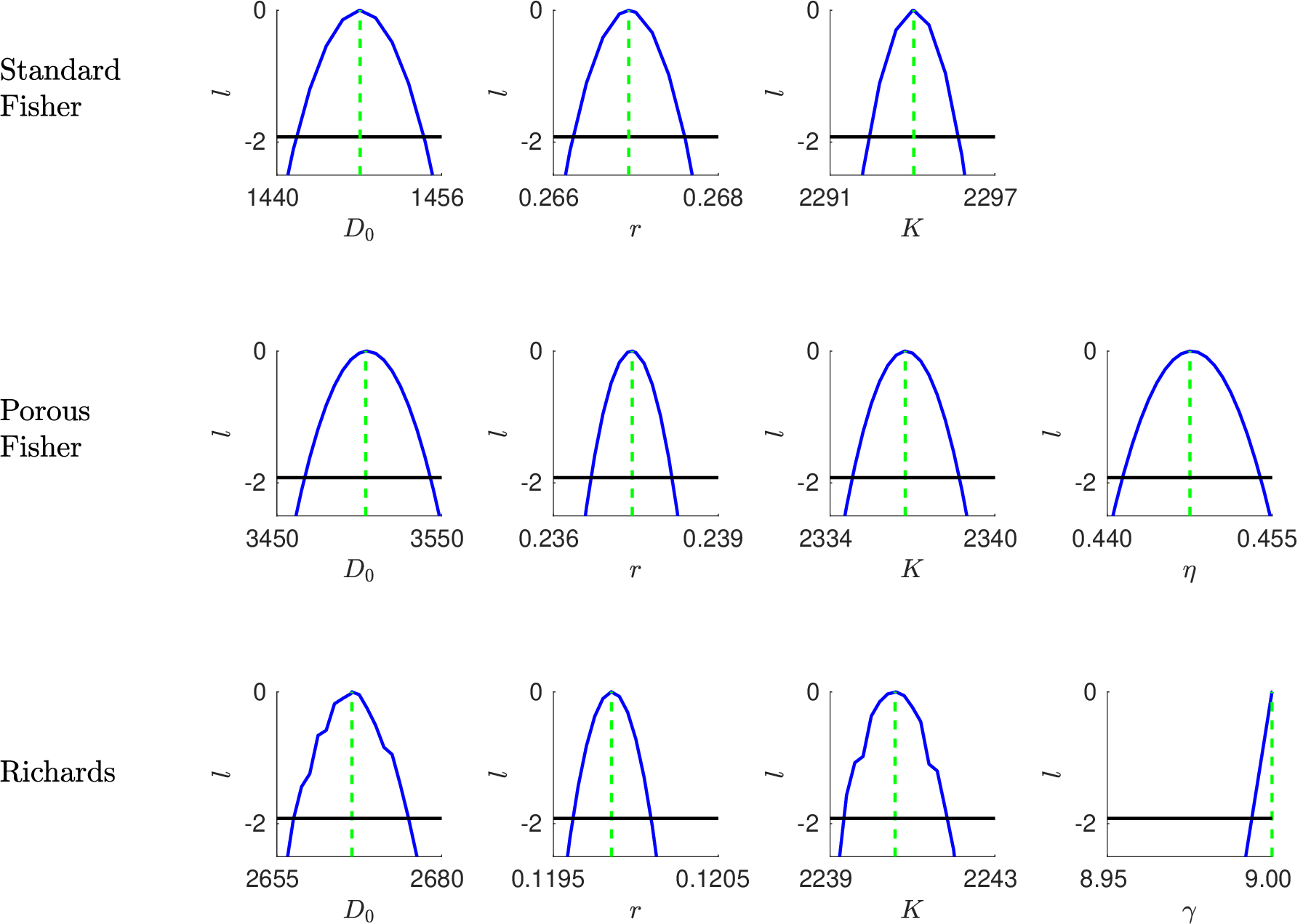}
    \caption{Profile likelihoods for Experiment 8, full dataset. }
\end{figure}


\newpage
\section{Profile likelihoods for down-sampled data}\label{apx:lowdata}

\begin{figure}[H]
    \centering
    \includegraphics[width=0.95\columnwidth]{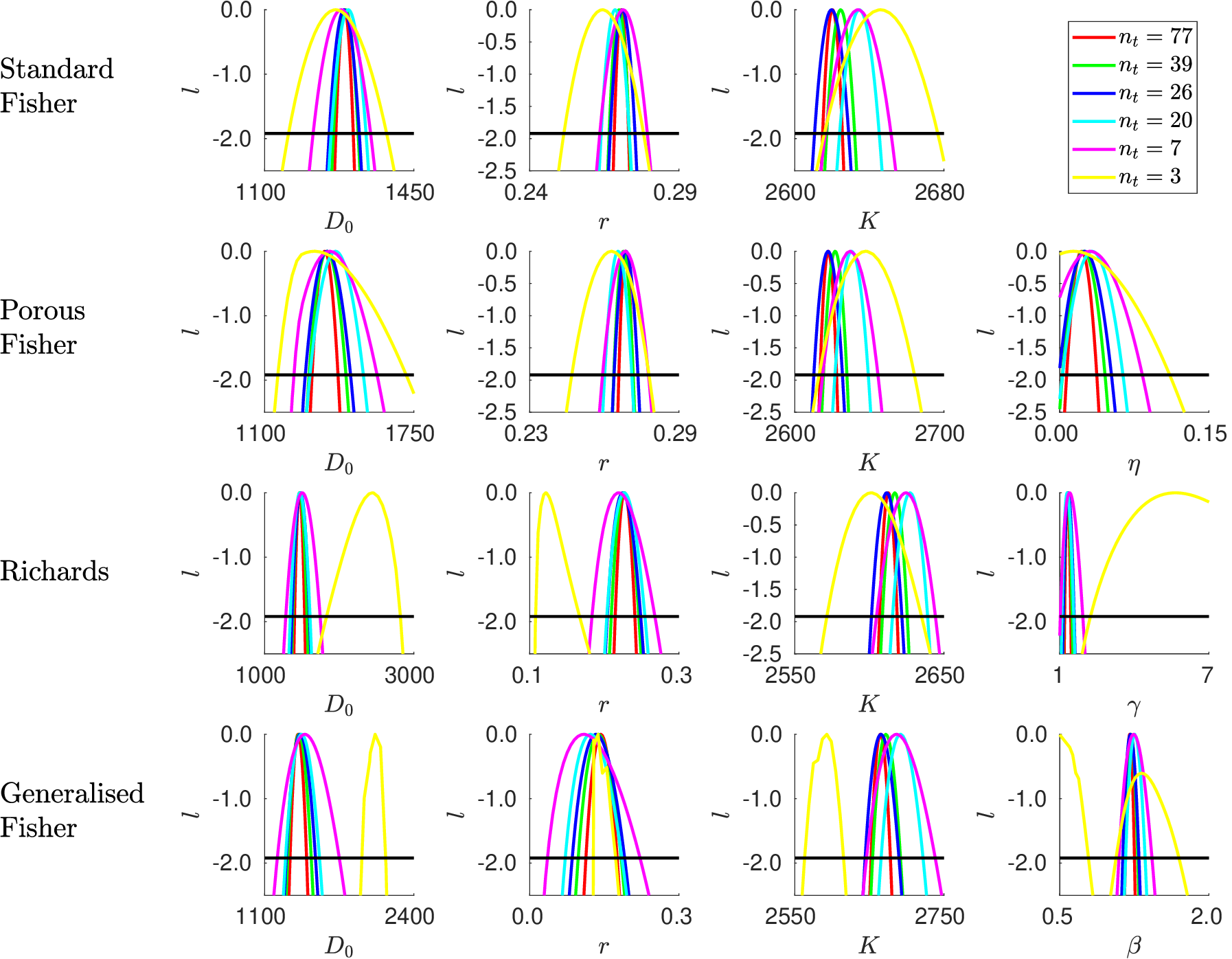}
    \caption{Profile likelihoods for the down-sampled datasets. A subset of these were presented in Fig.~\ref{fig:lowdata_pl_compare} of the main text.}
    \label{fig:lowdata_all}
\end{figure}

In Fig.~\ref{fig:lowdata_all} we present the profile likelihoods for the down-sampled datasets.


\newpage
\printbibliography

\end{document}